\documentclass[a4paper,11pt]{article}
\usepackage{amsfonts}
\usepackage{amsmath}
\usepackage{amssymb}
\usepackage{graphicx}
\usepackage{yfonts}
\usepackage{makeidx}
\usepackage{enumerate}
\usepackage{appendix}
\usepackage{verbatim}

\makeindex

\def \be {\begin{equation}}

\def \ee {\end{equation}}
\def \bea {\begin{eqnarray}}
\def \eea {\end{eqnarray}}
\def \nn {\nonumber}
\def \la {\langle}
\def \ra {\rangle}
\def \rr {\raise.35ex\hbox{\small $\prime$}\kern-.17em{\mbox{\large $\imath$}}}
\def \del {\partial}
\def \dels {\partial\kern-.6em /\kern.1em}
\def \As {{A\kern-.5em / \kern.5em}}
\def \Ds {D\kern-.7em / \kern.5em}

\def \a {\alpha}

\def \JJ {\mathfrak{J}}
\def \KK {\mathfrak{K}}
\def \MM {\mathfrak{M}}
\def \NN {\mathfrak{N}}
\def \VV {\mathfrak{V}}
\def \LL {{\cal L}}
\def \m {\mu}
\def \n {\nu}

\def \ks {k\kern-.5em /}
\def \lam {\lambda}
\def \Lam {\Lambda}

\def \om {\omega}

\def \vep {\varepsilon}

\begin{document}

\begin{titlepage}

    \thispagestyle{empty}
    \begin{flushright}
        \hfill{CERN-PH-TH/2012-213}\\
        \hfill{UCB-PTH-12/13}\\
    \end{flushright}

    \vspace{20pt}

    \begin{center}
        { \huge{\textbf{Freudenthal Gauge Theory }}}

        \vspace{30pt}

        {\large{{\bf Alessio Marrani$^{1}$, \ Cong-Xin Qiu$^{2}$, \ Sheng-Yu Darren Shih$^{3}$, \\  \vspace{5pt} \ Anthony Tagliaferro$^{3}$, and \ Bruno Zumino$^{3, 4}$
        }}}

        \vspace{20pt}

        {$1$ \it Theory division, CERN,\\
        CH 1211, Geneva 23, Switzerland\\
        \texttt{alessio.marrani@cern.ch}}

        \vspace{10pt}

        {$2$ \it School of Physics and Astronomy,\\
        University of Minnesota/Twin Cities,\\
        Minneapolis, MN 55455, USA\\
        \texttt{congxin.qiu@gmail.com}}

        \vspace{10pt}

        {$3$ \it Department of Physics and Center for Theoretical Sciences,\\University of California,\\
        Berkeley, CA 94720-7300, USA\\\texttt{atag@berkeley.edu}\\\texttt{s.y.darren.shih@berkeley.edu}}

          \vspace{10pt}

        $^4${ \sl Lawrence Berkeley National Laboratory, Theory Group, \\ Berkeley, CA 94720-8162, USA\\
\texttt{zumino@thsrv.lbl.gov}}


        \vspace{40pt}

        {ABSTRACT}
\end{center}

We present a novel gauge field theory, based on the \textit{Freudenthal
Triple System} (\textit{FTS}), a ternary algebra with mixed symmetry (not
completely symmetric) structure constants. The theory, named \textit{%
Freudenthal Gauge Theory} (\textit{FGT}), is invariant under two (off-shell)
symmetries: the gauge Lie algebra constructed from the \textit{%
FTS} triple product and a novel \textit{global} non-polynomial symmetry, the
so-called \textit{Freudenthal duality}.

Interestingly, a broad class of \textit{FGT} gauge algebras is provided by
the Lie algebras \textquotedblleft of type $\mathfrak{e}_{7}$" which occur
as conformal symmetries of Euclidean Jordan algebras of rank 3, and as
$U$-duality algebras of the corresponding (super)gravity theories in $D=4$.

We prove a \textit{No-Go Theorem}, stating the incompatibility of the
invariance under \textit{Freudenthal duality} and the coupling to
space-time vector \textit{and/or} spinor fields, thus forbidding non-trivial supersymmetric extensions of \textit{FGT}.

We also briefly discuss the relation between \textit{FTS} and the triple
systems occurring in BLG-type theories, in particular focusing on
superconformal Chern-Simons-matter gauge theories in $D=3$.

\end{titlepage}
\newpage \tableofcontents \newpage


\section{\label{Sec-Intro}Introduction}


The idea that a ternary algebra might be an essential structure of physical
theories has a long history.

In the early 70's, Nambu \cite{nambu} proposed a generalized Hamiltonian
system based on a ternary product, the \textit{Nambu-Poisson bracket}.
Despite some partial results (see \textit{e.g.} \cite{takhtajan} for a
comprehensive review), the quantization of the Nambu-Poisson bracket remains
a long-term puzzle.

However, ternary algebras and their applications to theoretical physics have
been object of intense study over the last four decades. The Jordan triple
product was exploited by G\"{u}naydin and G\"{u}rsey in their quest for a
formulation of quantum mechanics over different division algebras, including
octonions; this investigation led to the quadratic Jordan formulation of
quantum mechanics in terms of Jordan triple product \cite{G1,G2} that
extends to the octonionic quantum mechanics of \cite{G3}, which has no
formulation over an Hilbert space. Later on, a unified construction of Lie
algebras and Lie superalgebras over triple systems was achieved by Bars and G%
\"{u}naydin in \cite{G4}, and in \cite{G5} various composite models based on
ternary algebras were investigated. Two-dimensional superconformal algebras
over triple systems were then constructed in \cite{G6}; in particular,
Freudenthal triple systems were applied to $\mathcal{N}=4$ superconformal
algebras and gauged WZW models in \cite{G7}.

Recently, ternary algebras re-appeared in the study of $M$-theory by Bagger
and Lambert \cite{BLG} and by Gustavsson \cite{gustavsson}, in which a
ternary Lie-$3$ algebra is proposed as the underlying gauge symmetry
structure on a stack of supersymmetric $M2$-branes; this is the famous BLG
theory (for a recent review and list of Refs., see \textit{e.g.} \cite%
{Lambert-rev}). When taking the Nambu-Poisson bracket as an
infinite-dimensional generalization of the Lie-$3$ bracket, one gets from
the BLG theory a novel six-dimensional field theory, which can be
interpreted as a non-commutative version of the $M5$-brane theory \cite{ho}%
.\medskip

In the present paper, we propose a novel gauge field theory, based on
another ternary algebra: the \textit{Freudenthal Triple System}\footnote{%
Historically, there are several different notions of \textit{Freudenthal
Triple System}, which differ by the symmetry structure of their triple
product. They were introduced in mathematics in order to address different
algebraic properties of the triple system. Although simply related,
different definitions of \textit{FTS} have different properties, which of
course can be translated from one to another. In the physics literature, the
\textit{FTS} we focus on in this paper is sometimes also called \textit{%
generalized Freudenthal Triple System}, which makes the derivation property
more transparent.
\par
Since there is no general agreement on the definition, we will simple denote
the triple system in this paper by \textit{Freudenthal Triple System} (%
\textit{FTS}). The \textit{FTS} introduced in $\mathcal{N}=2$
Maxwell-Einstein supergravity and its $\mathcal{N}>2$ generalizations \cite%
{Gunaydin:1983rk, Gunaydin:1983bi} (see also \textit{e.g.} \cite{G-Lects}
and \cite{Borsten:2011ai, Borsten:2011nq} for recent reviews) can be
regarded as some special cases of it.} (\textit{FTS}). We call this theory
\textit{\textquotedblleft Freudenthal Gauge Theory"} (FGT). In its simplest
setup, FGT contains a bosonic scalar field $\phi (x)$ valued in the \textit{%
FTS} $\KK$ together with a gauge field $A_{\m}(x)$ taking values in the
symmetric product $\KK\otimes _{s}\KK$. Similar to the BLG theory, the gauge
transformation is constructed from a triple product defined over the \textit{%
FTS} $\KK$. However, unlike the totally anti-symmetric Lie-$3$ bracket used
in the BLG theory, in general the \textit{FTS} triple product does not have
a simple symmetry structure with respect to the exchange of a pair of its
arguments. Nevertheless, one can still prove that the gauge invariance of
FGT is guaranteed by the algebraic properties of the \textit{FTS}.

Besides the \textit{off-shell} gauge symmetry, FGT also possesses a novel
\textit{global (off-shell)} symmetry, the so-called \textit{Freudenthal
duality} (\textit{F-duality}). This is a non-linear, non-polynomial mapping
from $\KK$ to $\KK$, relying on non-linear identities which can be traced
back to the early days of the mathematical investigation of \textit{FTS}'s
\cite{brown}. The name \textit{Freudenthal duality} is much more recent, and
it was introduced within physical literature in \cite{duff}, in the study of
Maxwell-Einstein supergravity theories (MESGT's) in $D=4$ space-time
dimensions based on symmetric scalar manifolds and with \textit{%
non-degenerate} groups \textit{of type }$E_{7}$ \cite%
{brown,duff,alessio,Ceresole:2011xd,Ferrara:2011aa,Ferrara:2011dz,Ferrara:2012qp}
as generalized electric-magnetic ($U$-)duality\footnote{%
Here $U$-duality is referred to as the \textquotedblleft
continuous\textquotedblright\ symmetries of \cite{CJ-1}. Their discrete
versions are the $U$-duality non-perturbative string theory symmetries
introduced in \cite{HT-1}.} symmetries. In such a framework, \textit{%
F-duality} was observed as a non-polynomial, anti-involutive mapping on $\KK$%
-valued black hole charges \textit{(i.e.} fluxes of the Abelian $2$-form
field strengths) which keeps the Bekenstein-Hawking \cite{BH-1, BH-2} black
hole entropy invariant \cite{duff}. Further generalization to a generic $%
\mathcal{N}=2$ special K\"{a}hler geometry, to its $\mathcal{N}>2$
generalization and to the so-called effective black hole potential governing
the scalar flows has been discussed in \cite{alessio}.

At any rate, FGT, in its simplest setup presented in this paper, can be
regarded as the simplest gauge theory admitting \textit{F-duality} as
\textit{global} symmetry. Despite the \textit{No-Go theorem} proved in Sec. %
\ref{Subsec-No--Go}, a slight generalization of the FGT will be presented in
a companion paper \cite{MST-2}.

Intriguingly, as discussed in\ Sec. \ref{FGT-SC}, FGT shares the same
symmetry structures as the \textquotedblleft quaternionic level" of
Faulkner's construction \cite{Faulkner-constr}, which relates triple systems
to pairs $(\mathfrak{g},\mathbf{V})$ of a metric Lie algebra $\mathfrak{g}$
and a suitable representation $\mathbf{V}$. After the treatment \cite%
{F-1,F-2}, an interesting similarity between FGT and the bosonic sector of $%
\mathcal{N}=3$, $D=3$ superconformal (SC) Chern-Simons-matter (CSM) gauge
theories can be envisaged. An important difference relies in supersymmetry,
which in FGT, as discussed in Sec. \ref{Sec-Gen}, is essentially spoiled by
the enforcement of \textit{global} invariance under \textit{F-duality}; this
affects also other terms in the Lagrangian, \textit{e.g.} the scalar
potential (\textit{quartic} in FGT, \textit{sextic} in BLG-type theories).

All in all, we can observe that, with some important differences pointed out
along the present investigation, the same symmetry structures are shared
(with different implementations and physical meanings) by three (\textit{a
priori} very different) classes of theories, namely : ($D=3$) FGT (\textit{%
non-supersymmetric}), $D=4$ MESGT (with various amounts of \textit{local}
supersymmetry) and $D=3$ SC CSM gauge theory (with $\mathcal{N}=3$ \textit{%
global} supersymmetry). Further details and results will be reported in a
companion paper \cite{MST-2}. \bigskip\ \

This paper is organized as follows.

We start by recalling the relation between \textit{FTS}, rank-$3$ Euclidean
Jordan algebras and exceptional Lie algebras (Sec. \ref{Subsec-Lie-Algebras}%
); the treatment is then generalized in Sec. \ref{Subsec-General-Case}. The
axiomatic definition of a \textit{FTS} and the general symmetry of its
structure constants are then discussed in Secs. \ref{Subsec-FTS-def} and \ref%
{Subsec-f_abcd}. The \textit{Freudenthal duality} for a generic \textit{FTS}
is introduced in Sec. \ref{Subsec-FD-def}, along with a discussion of its
basic properties.

The global transformation constructed from the \textit{FTS} triple product
is introduced in Sec. \ref{Subsec-Global}, and its gauging is discussed in
Sec. \ref{Subsec-Gauge}. Then, in Sec. \ref{Subsec-Minimal} we propose a
bosonic Lagrangian density that exhibits both \textit{FTS} gauge
transformations and (global) \textit{F-duality} as \textit{off-shell}
symmetries, and we provide a detailed proof of its invariance under such
symmetries. The class of FGT gauge Lie algebras \textit{of type} $\mathfrak{e%
}_{7}$ is considered in Sec. \ref{Gauge-Algebras-type-e7}, and the
intriguing relation between the corresponding FGT and $D=4$ MESGT's with $U$%
-duality symmetry given by such Lie algebras \textit{of type} $\mathfrak{e}%
_{7}$ is discussed in Sec. \ref{FGT-sugra}.

The possible generalization of the simplest FGT Lagrangian introduced in
Sec. \ref{Subsec-Minimal} is discussed in Sec. \ref{Sec-Gen}, in which the
\textit{FTS} $\KK$ is coupled to the most general algebraic system, and the
mathematical structure required for a consistent definition of \textit{%
F-duality} is investigated (Sec. \ref{Subsec-Couple}); a \textit{No-Go
theorem} is proved in Sec. \ref{Subsec-No--Go}.

The intriguing similarities (and important differences) between FGT and (the
bosonic sector of) $\mathcal{N}=3$ SC CSM gauge theories in $D=3$ are
discussed in Sec. \ref{FGT-SC}.

The concluding Sec. \ref{Sec-Conclusion} contains a summary, along with some
remarks and an outlook of further developments.

Three Appendices conclude the paper. Apps. \ref{App-FD} and \ref%
{App-Symm-Scalar-Kinetic} respectively contain details on the \textit{%
F-duality} and on the FGT scalar kinetic term, whereas App. \ref%
{App-Ind-Axioms} lists the induced axioms needed for the discussion of the
generalization of FGT and in the proof of the \textit{No-Go theorem} of Sec. %
\ref{Subsec-No--Go}.

As mentioned above, further results and more detailed analysis of some
topics mentioned along the paper will be reported in a companion work \cite%
{MST-2}.


\section{\label{Sec-FTS}Freudenthal Triple Systems (\textit{FTS}'s)}

\subsection{\label{Subsec-Lie-Algebras}Rank-$3$ Jordan Algebras and Lie
Algebras}

The \textit{Freudenthal Triple System} (\textit{FTS}) $\KK$ was first
introduced by Freudenthal in his study of exceptional Lie algebras \cite%
{freudenthal,freudenthal-FTS-1,freudenthal-FTS-2} (see also \cite{Faulkner}%
). In the original construction, $\KK$ is defined to be the direct sum of
two copies of a \textit{Jordan Triple System} (\textit{JTS}) $\JJ$ and two
copies of real numbers\footnote{%
Namely, the ground field was chosen to be $\mathbb{R}$. Other choices are of
course possible (such as $\mathbb{Z}$ or $\mathbb{C}$), but we will not deal
with them in the present investigation.} $\mathbb{R}$:%
\begin{equation}
\KK(\JJ)\equiv \JJ\oplus \JJ\oplus \mathbb{R}\oplus \mathbb{R}.  \label{FTS}
\end{equation}%
Over the vector space $\KK(\JJ)$, one can introduce a \textit{symplectic}
invariant $2$-form, as well as a \textit{triple product}. The latter is
defined via the completely symmetric tri-linear form (also known as \textit{%
cubic norm}) of the \textit{JTS }$\JJ$, and it can be re-interpreted as a
linear map $\LL_{\phi _{I}\phi _{J}}$ over $\KK$ parametrized by a pair of
elements $\phi _{I},\phi _{J}\in \KK$ (\textit{cfr.} definition (\ref{LL-def}%
)).\medskip

In Freudenthal's construction of exceptional Lie algebras, the \textit{JTS} $%
\JJ$ is restricted to a rank-$3$ \textit{simple} Euclidean \textit{Jordan
algebra }$\widehat{\mathfrak{J}}$, namely $\widehat{\mathfrak{J}}=\mathbb{R}$
or $\widehat{\mathfrak{J}}=J_{3}^{\mathfrak{A}}\equiv H_{3}(\mathfrak{A})$,
where $H_{3}(\mathfrak{A})$ stands for the algebra of Hermitian $3\times 3$
matrices with entries taking values in one of the four \textit{normed
division algebras} $\mathfrak{A}=\mathbb{R}$ (real numbers), $\mathbb{C}$
(complex numbers), $\mathbb{H}$ (quaternions), $\mathbb{O}$ (octonions) (see
\textit{e.g.} \cite{McCrimmon}). Then, by introducing in $\KK(\widehat{%
\mathfrak{J}})$ the submanifold
\begin{equation}
\MM_{\widehat{\mathfrak{J}}}\equiv \big\{\phi _{I}\in \KK(\widehat{\mathfrak{%
J}})\,|\,\LL_{\phi _{I}\phi _{I}}\phi _{J}=0,\;\forall \,\phi _{J}\in \KK(%
\widehat{\mathfrak{J}})\big\},  \label{MM}
\end{equation}%
the five \textit{exceptional} (finite-dimensional) Lie algebras $\mathfrak{G}%
=\mathfrak{g}_{2},\mathfrak{f}_{4},\mathfrak{e}_{6},\mathfrak{e}_{7},%
\mathfrak{e}_{8}$ arise as the the direct sum of the algebra $\mbox{Inv}(\MM%
_{\widehat{\mathfrak{J}}})$ that keeps $\MM_{\widehat{\mathfrak{J}}}$
invariant, together with a copy of $\mathfrak{su}(2)$ and two copies
(namely, an $\mathfrak{su}(2)$-doublet) of $\KK(\widehat{\mathfrak{J}})$
\cite{freudenthal,yamaguti}:%
\begin{equation}
\mathfrak{G}=\mbox{Inv}(\MM_{\widehat{\mathfrak{J}}})\oplus \mathfrak{su}%
(2)\oplus \KK(\widehat{\mathfrak{J}})\oplus \KK(\widehat{\mathfrak{J}}).
\label{exc-Lie}
\end{equation}

As a vector space, $\KK\left( \widehat{\mathfrak{J}}\right) $ may be
regarded as the \textit{representation space} of a non-trivial\footnote{%
Such a representation is not necessarily the smallest one. A counter-example
is provided \textit{e.g.} by $\mathfrak{sp}(6)=\mbox{Inv}(\MM_{J_{3}^{%
\mathbb{R}}})$, whose smallest non-trivial symplectic irrep. is the
fundamental $\mathbf{6}$. However, $\KK(J_{3}^{\mathbb{R}})$ has dimension $%
14$, and it is based on the rank-$3$ completely antisymmetric irrep. $%
\mathbf{14}^{\prime }$, which exhibits a \textit{completely symmetric} rank-$%
4$ invariant structure.
\par
However, a suitable \textit{FTS} $\KK$ on the $\mathbf{6}$ can also be
constructed; see point 2 in Sec. \ref{FGT-SC}.} \textit{symplectic}
representation $\mathbf{R}$ of the algebra $\mbox{Inv}(\MM_{\widehat{%
\mathfrak{J}}})$ itself, introduced in (\ref{exc-Lie}):%
\begin{equation}
\KK\left( \widehat{\mathfrak{J}}\right) \sim \mathbf{R}\left( \mbox{Inv}(\MM%
_{\widehat{\mathfrak{J}}})\right) .  \label{FTS-R}
\end{equation}%
\textit{At least} for $\mathbf{R}$ irreducible, $\mbox{Inv}(\MM_{\widehat{%
\mathfrak{J}}})$ is maximally (and non-symmetrically) embedded into the
symplectic algebra $\mathfrak{sp}\left( \KK\left( \widehat{\mathfrak{J}}%
\right) \right) $ through the \textit{Gaillard-Zumino (GZ) embedding }\cite%
{GZ} (see also \textit{e.g.} \cite{AGZ-rev} for a recent review)%
\begin{equation}
\begin{array}{l}
\mathfrak{sp}\left( \KK\left( \widehat{\mathfrak{J}}\right) \right) \supset %
\mbox{Inv}(\MM_{\widehat{\mathfrak{J}}}); \\
\\
\mathbf{Fund}\left( \mathfrak{sp}\left( \KK\left( \widehat{\mathfrak{J}}%
\right) \right) \right) =\mathbf{R}\left( \mbox{Inv}(\MM_{\widehat{\mathfrak{%
J}}})\right) .%
\end{array}
\label{GZ-pre}
\end{equation}%
This can be regarded as a consequence of the following \textit{Theorem} by
Dynkin (Th. 1.5 of \cite{dynkin-2}, more recently discussed \textit{e.g.} in
\cite{Lorente}) : Every \textit{irreducible} group of unimodular linear
transformations of the N-dimensional complex space (namely, a group of
transformations which does not leave invariant a proper subspace of such a
space) is maximal either in $SL(N)$ (if the group does not have a bilinear
invariant), or in $Sp(N)$ (if it has a skew-symmetric bilinear invariant),
or in $O(N)$ (if it has a symmetric bilinear invariant). Exceptions to this
rule are listed in Table VII of \cite{Lorente}.

For later convenience, we introduce the number $f$ as (\textit{cfr.} (\ref%
{FTS-R}))%
\begin{equation}
\text{dim}_{\mathbb{R}}\mathbf{Fund}\left( \mathfrak{sp}\left( \KK\left(
\widehat{\mathfrak{J}}\right) \right) \right) =\text{dim}_{\mathbb{R}}%
\mathbf{R}\left( \mbox{Inv}(\MM_{\widehat{\mathfrak{J}}})\right) =\text{dim}%
_{\mathbb{R}}\KK\left( \widehat{\mathfrak{J}}\right) \equiv f,
\label{dim-f-def}
\end{equation}%
which is even whenever the symplectic $2$-form on $\KK\left( \widehat{%
\mathfrak{J}}\right) $ is \textit{non-degenerate} (as we will assume
throughout).

From (\ref{exc-Lie}) and (\ref{GZ-pre}), it thus follows that the invariance
subalgebra $\mbox{Inv}(\MM_{\widehat{\mathfrak{J}}})$ can be equivalently
defined as the intersection of two Lie algebras : the \textit{symplectic}
one $\mathfrak{sp}\left( \KK\left( \widehat{\mathfrak{J}}\right) \right) $
in (\ref{GZ-pre}) and the \textit{exceptional} one $\mathfrak{G}$($=%
\mathfrak{g}_{2},\mathfrak{f}_{4},\mathfrak{e}_{6},\mathfrak{e}_{7},%
\mathfrak{e}_{8}$) in (\ref{exc-Lie}):%
\begin{equation}
\mbox{Inv}(\MM_{\widehat{\mathfrak{J}}})=\mathfrak{sp}\left( \KK\left(
\widehat{\mathfrak{J}}\right) \right) \cap \mathfrak{G}.  \label{inters}
\end{equation}

\subsection{\label{Subsec-General-Case}General Case}

Within Freudenthal's formulation, the above construction can be repeated for
a \textit{generic} FTS $\KK$ , by generalizing (\ref{MM}) to the submanifold%
\begin{equation}
\MM_{\mathfrak{J}}\equiv \big\{\phi _{I}\in \KK(\mathfrak{J})\,|\,\LL_{\phi
_{I}\phi _{I}}\phi _{J}=0,\;\forall \,\phi _{J}\in \KK(\mathfrak{J})\big\},
\label{MM-2}
\end{equation}%
and thus introducing its invariance algebra $\mbox{Inv}(\MM_{\mathfrak{J}})$.

It is however worth remarking that, in this general case, neither $\mbox{Inv}%
(\MM_{\mathfrak{J}})$ nor%
\begin{equation}
\mathfrak{G}=\mbox{Inv}(\MM_{\mathfrak{J}})\oplus \mathfrak{su}(2)\oplus \KK(%
\mathfrak{J})\oplus \KK(\mathfrak{J})  \label{exc-Lie-gen}
\end{equation}%
(this latter generalizing (\ref{exc-Lie}) to a generic \textit{JTS} $%
\mathfrak{J}$), along with their possible non-compact real forms, are
necessarily simple.

Nonetheless, it still holds that, as a vector space, $\KK\left( \mathfrak{J}%
\right) $ may be regarded as the \textit{representation space} of the
relevant \textit{symplectic} representation $\mathbf{R}$ of the invariance
subalgebra $\mbox{Inv}(\MM_{\mathfrak{J}})$ of $\MM_{\mathfrak{J}}$ (\ref%
{MM-2}):%
\begin{equation}
\KK\left( \mathfrak{J}\right) \sim \mathbf{R}\left( \mbox{Inv}(\MM_{%
\mathfrak{J}})\right) .  \label{FTS-R-gen}
\end{equation}%
\medskip\

Before proceeding to analyze the axiomatic definition of \textit{FTS}, we
remark that, as mentioned in Footnote 1, in the mathematics literature there
are several different notions of \textit{FTS}, which differ by the symmetry
structure of the corresponding triple product (see for instance \cite%
{brown,Faulkner,ferrar}). All of these \textquotedblleft \textit{FTS}%
's\textquotedblright\ are closely inter-related by simple redefinitions;
however, because they exhibit different symmetry properties, some algebraic
properties of the \textit{FTS} are manifest only within a specific
formulation.

\subsection{\label{Subsec-FTS-def}Axiomatic Definition}

We define an \textit{FTS} to be a particular \textit{Symplectic Triple System%
} \cite{Okubo-1,Okubo-2}, which is a symplectic vector space $\KK$ equipped
with a (not necessarily completely symmetric) \textit{triple product}%
\begin{equation}
T:\left\{
\begin{array}{l}
\KK\otimes \KK\otimes \KK\rightarrow \KK; \\
\\
\phi _{I},\phi _{J},\phi _{K}\mapsto T\left( \phi _{I},\phi _{J},\phi
_{K}\right) .%
\end{array}%
\right.  \label{T-def}
\end{equation}%
In the following, for brevity's sake, we will denote $T\left( \phi _{I},\phi
_{J},\phi _{K}\right) \equiv \phi _{I}\phi _{J}\phi _{K}$.

By introducing the symplectic form as\footnote{%
Subscripts \textquotedblleft $s$" and \textquotedblleft $a$" respectively
stand for \textit{symmetric} and \textit{antisymmetric}.}%
\begin{equation}
\la\cdot ,\cdot \ra:\left\{
\begin{array}{l}
\KK\otimes _{a}\KK\rightarrow \mathbb{R}; \\
\\
\phi _{I},\phi _{J}\mapsto \la\phi _{I},\phi _{J}\ra,%
\end{array}%
\right.  \label{sympl-def}
\end{equation}%
in an \textit{FTS} the triple product (\ref{T-def}) satisfies the following
\textit{axioms}:

\begin{itemize}
\item[($i$)] $\phi _{I}\phi _{J}\phi _{K}=\phi _{J}\phi _{I}\phi _{K};$

\item[($ii$)] $\phi _{I}\phi _{J}\phi _{K}=\phi _{I}\phi _{K}\phi
_{J}+2\lambda \,\la\phi _{J},\phi _{K}\ra\phi _{I}+\lambda \,\la\phi
_{I},\phi _{K}\ra\phi _{J}-\lambda \,\la\phi _{I},\phi _{J}\ra\phi _{K};$

\item[($iii$)] $\phi _{L}\phi _{M}(\phi _{I}\phi _{J}\phi _{K})=(\phi
_{L}\phi _{M}\phi _{I})\phi _{J}\phi _{K}+\phi _{I}(\phi _{L}\phi _{M}\phi
_{J})\phi _{K}+\phi _{I}\phi _{J}(\phi _{L}\phi _{M}\phi _{K});$

\item[($iv$)] $\la\phi _{L}\phi _{M}\phi _{I},\phi _{J}\ra+\la\phi _{I},\phi
_{L}\phi _{M}\phi _{J}\ra=0,$
\end{itemize}

where $\lambda $ is an arbitrary (real) constant\footnote{%
Axioms ($i$)-($iv$) define the most general \textit{FTS }$\KK$, which does
not necessarily enjoys the decomposition (\ref{FTS}) in terms of an
underlying \textit{JTS} $\JJ$ (as in the original Freudenthal's
construction).
\par
A counterexample is provided by Example 1 of \cite{Faulkner}, in which $%
\mathfrak{g}=\mathfrak{sp}\left( 2l\right) $. In $\mathcal{N}=1$, $D=4$
supergravity, this corresponds to a theory in which the scalar fields
parametrize the \textit{upper Siegel half-plane}; see \textit{e.g.} a recent
treatment in \cite{ADFT-N=1-attractors}.}.

By introducing, for any pair $\phi _{L},\phi _{M}\in \KK$, a linear operator
$\LL_{\phi _{L}\phi _{M}}\in \mathfrak{gl}(\KK)$ acting on $\phi _{K}\in \KK$
as%
\begin{equation}
\LL_{\phi _{I}\phi _{J}}:\left\{
\begin{array}{l}
\KK\otimes _{s}\KK\rightarrow \KK; \\
\\
\phi _{I},\phi _{J}\mapsto \LL_{\phi _{I}\phi _{J}}\,\phi _{K}\equiv \phi
_{I}\phi _{J}\phi _{K},%
\end{array}%
\right.  \label{LL-def}
\end{equation}%
axiom ($iii$) yields that\ $\LL_{\phi _{I}\phi _{J}}$ is a \textit{derivation%
} with respect to the \textit{FTS} triple product $T$ (\ref{T-def}).

On the other hand, axiom ($i$) implies%
\begin{equation}
\LL_{\phi _{I}\phi _{J}}=\LL_{\phi _{J}\phi _{I}},  \label{symm-L-call}
\end{equation}%
which justifies the \textit{symmetric} tensor product of $\KK$'s in the
definition (\ref{LL-def}) itself.

By virtue of the definition (\ref{LL-def}), one can reformulate axioms ($iii$%
) and ($iv$) as follows:

\begin{itemize}
\item[($iii^{\prime }$)] $\LL_{\phi _{L}\phi _{M}}\,(\phi _{I}\phi _{J}\phi
_{K})=(\LL_{\phi _{L}\phi _{M}}\,\phi _{I})\phi _{J}\phi _{K}+\phi _{I}(\LL%
_{\phi _{L}\phi _{M}}\,\phi _{J})\phi _{K}+\phi _{I}\phi _{J}(\LL_{\phi
_{L}\phi _{M}}\,\phi _{K});$

\item[($iv^{\prime }$)] $\LL_{\phi _{L}\phi _{M}}\,\la\phi _{I},\phi _{J}\ra=%
\la\LL_{\phi _{L}\phi _{M}}\,\phi _{I},\phi _{J}\ra+\la\phi _{I},\LL_{\phi
_{L}\phi _{M}}\,\phi _{J}\ra=0.$
\end{itemize}

In particular, the reformulation ($iv^{\prime }$) of axiom ($iv$) makes
manifest the fact the symplectic form $\la\cdot ,\cdot \ra$ (\ref{sympl-def}%
) is \textit{invariant} under $\LL_{\phi _{I}\phi _{J}}$. Thus, $\LL_{\phi
_{I}\phi _{J}}$ is valued in a certain Lie algebra $\mathfrak{g}$, which
exhibits a symplectic bilinear invariant structure in the relevant
representation $\mathbf{R}$ to which $\phi _{I}$ belongs. \textit{At least}
when such a representation space is \textit{irreducible}, through the GZ
embedding \cite{GZ}, or equivalently through the abovementioned Dynkin
Theorem \cite{dynkin-2}%
\begin{equation}
\mathfrak{g}\overset{GZ}{\subset }\mathfrak{sp}(\KK)\subset \mathfrak{gl}(\KK%
):\mathbf{R}\left( \mathfrak{g}\right) =\mathbf{Fund}\left( \mathfrak{sp}%
\right) =\mathbf{Fund}\left( \mathfrak{gl}\right) ,  \label{GZ-g}
\end{equation}%
one has%
\begin{equation}
\LL_{\phi _{I}\phi _{J}}\in \mathfrak{g}\overset{GZ}{\subset }\mathfrak{sp}(%
\KK)\subset \mathfrak{gl}(\KK).  \label{LL-sympl-valued}
\end{equation}%
\medskip Within Freudenthal's construction, an important class of algebras
is given by $\mathfrak{g}=\mbox{Inv}(\MM_{\widehat{\mathfrak{J}}})$
introduced above. The Lie algebra $\mathfrak{g}$ will be identified below as
the gauge Lie algebra of the \textit{Freudenthal gauge theory}.

It is worth remarking here that for $\lam\neq 0$ axiom ($iv$) can actually
be derived from axioms ($i$)-($iii$). Mathematically, whenever $\lam\neq 0$
axiom ($ii$) yields a compatibility condition that constrains the structure
of the triple product (\ref{T-def}) and the symplectic form (\ref{sympl-def}%
), and hence the non-trivial algebraic structure of the \textit{FTS }itself.
We anticipate that axiom ($iii$) can be regarded as the \textit{%
\textquotedblleft FTS counterpart"} of the so-called \textit{%
\textquotedblleft fundamental identity"} of Lie-$3$ algebras (see Sec. \ref%
{FGT-SC}). On the other hand, for $\lam=0$ axioms ($i$)-($iii$) reduce to
the defining properties of a Lie-$3$ algebra over Grassmannian numbers,
which in general is \textit{not} a \textit{FTS}. And hence, in order to
restore the algebraic structure of the \textit{FTS} $\KK$, one has to
further impose axiom ($iv$) as a compatibility condition between the (now
totally symmetric) triple product (\ref{T-def}) and the symplectic form (\ref%
{sympl-def}).

At any rate, in the present investigation we regard an \textit{FTS} $\KK$ as
a \textit{Symplectic Triple System} \cite{Okubo-1,Okubo-2} with $\lam\neq 0$%
, and we include ($iv$) (or equivalently ($iv^{\prime }$)) as part of the
defining axioms, so that the most generic situation will be considered.

\subsection{\label{Subsec-f_abcd}\textit{FTS} Structure Constants and their
Invariance}

In order to make our treatment more explicit yet basis-dependent, it is
convenient to introduce a basis $\{e_{a}\}$ of $\KK$, such that $\phi =\phi
^{a}e_{a}$ ($a=1,...,f$; $f=\dim_\mathbb{R}(\mathfrak{K})$, (\ref{dim-f-def}%
)). Thus, one can define the \textit{symplectic metric} $\om_{ab}$ and the
\textit{FTS} \textit{(triple product) structure constants} ${f_{abc}}^{d}$
respectively as%
\begin{equation}
\la e_{a},e_{b}\ra\equiv \om_{ab}\nn=-\om_{ba};  \label{omega-def}
\end{equation}%
\begin{equation}
e_{a}e_{b}e_{c}\equiv {f_{abc}}^{d}e_{d}.  \label{f-def-2}
\end{equation}%
As mentioned above, $\omega _{ab}$ is \textit{invariant} under $\mathfrak{g}$
(recall (\ref{GZ-g}) and (\ref{LL-sympl-valued})). Furthermore, when $\omega
_{ab}$ is \textit{non-degenerate} (which we will always assume to hold true
in this paper), an isomorphism is defined between the vector space $\KK$ and
its dual space, and hence one can lower\footnote{%
We adopt the NE-WS convention when raising or lowering the indices using the
symplectic metric.} the last index of the \textit{FTS} structure constants
as follows:%
\begin{equation}
f_{abcd}\equiv {f_{abc}}^{e}\om_{ed}.  \label{lower-sympl}
\end{equation}%
By virtue of definitions (\ref{omega-def}), the defining axioms ($i$)-($iv$)
of the \textit{FTS} $\KK$ can be rewritten as follows:

\begin{itemize}
\item[($i$)] $f_{abcd}=f_{bacd};$

\item[($ii$)] $f_{abcd}=f_{acbd}+2\lambda \omega _{ad}\omega _{bc}-\lambda
\omega _{ca}\omega _{bd}-\lambda \omega _{ab}\omega _{cd};$

\item[($iii$)] $f_{abc}^{\phantom{abc}d}f_{efd}^{\phantom{efd}g}=f_{efc}^{%
\phantom{efa}d}f_{abd}^{\phantom{dbc}g}+f_{ecf}^{\phantom{efb}d}f_{adb}^{%
\phantom{adc}g}+f_{fce}^{\phantom{efc}d}f_{bda}^{\phantom{abd}g};$

\item[($iv$)] $f_{abcd}=f_{abdc}.$
\end{itemize}

It is worth stressing here that the non-complete symmetry of the \textit{FTS}
triple product $T$ (\ref{T-def}) (as yielded by axioms ($i$) and ($ii$))
implies the non-complete symmetry of the\textit{\ }rank-$4$ tensor of
\textit{FTS} structure constants $f_{abcd}$ (\ref{lower-sympl}). However,
note that axioms ($i$), ($ii$), and ($iv$) imply the \textit{structure
constants} to be symmetric also under exchange of the first and last pair of
its indices:%
\begin{equation}
f_{abcd}=f_{cdab},  \label{symm-f}
\end{equation}%
a property which will be important in the construction of a Chern-Simons
action for the gauge fields of the \textit{\textquotedblleft Freudenthal
gauge theory"} (see next Sections).

Summarizing, the general symmetry properties of $f_{abcd}$, as implied by
axioms ($i$), ($ii$) and ($iv$), are given by%
\begin{equation}
f_{abcd}=f_{\left( \left( ab\right) ,\left( cd\right) \right) }.
\label{symm-f-gen}
\end{equation}%
${f_{abc}}^{d}$ and ${f_{abcd}}$ are rank-$4$ invariant tensors of the Lie
algebra $\mathfrak{g}$ (\ref{GZ-g})-(\ref{LL-sympl-valued}). Under certain
further restrictions (see point 2 in Sec. \ref{FGT-SC}), the symmetry can be
extended to $\mathfrak{sp}(\KK)$ itself. It is here worth recalling that
Kantor gave a complete classification of the finite dimensional triple
systems that can arise in Lie algebras \cite{G11} (see also \cite{Palmkvist}%
); in particular, Kantor and Skopets showed that there is a one-to-one
correspondence between simple Lie algebras and simple \textit{FTS}'s with a
non-degenerate bilinear form \cite{G12}.

\subsection{\label{Subsec-FD-def}Freudenthal Duality}

Whenever the completely symmetric part of $f_{abcd}$ is non-vanishing, from
the definition of the \textit{FTS} triple product (\ref{T-def}) and of the
symplectic form (\ref{sympl-def}) one can define a \textit{quartic} $%
\mathfrak{g}$-invariant structure $\Delta (\phi )$ for any $\phi \in \KK$,
as follows\footnote{%
Even if here $f_{abcd}$ is not (necessarily) completely symmetric in the
present framework, we adopt the same normalization of \cite{duff} and \cite%
{alessio}.} (\textit{cfr.} (25c) of \cite{duff}; $T(\phi) \equiv \phi \phi
\phi$):%
\begin{equation}
\Delta :\left\{
\begin{array}{l}
\KK\rightarrow \mathbb{R}; \\
\\
\phi \mapsto \Delta (\phi )\equiv \frac{1}{2}\la\phi \phi \phi ,\phi \ra=%
\frac{1}{2}f_{abcd}\phi ^{a}\phi ^{b}\phi ^{c}\phi ^{d}.%
\end{array}%
\right.  \label{Delta-def}
\end{equation}

Such a quartic form has appeared in physical literature \textit{e.g.} in the
formula for the Bekenstein-Hawking \cite{BH-1,BH-2} entropy of spherically
symmetric, asymptotically flat, static, extremal black hole solutions of $%
D=4 $ supergravity theories whose $U$-duality Lie algebra is a particular
non-compact, real form of $\mbox{Inv}(\MM_{\widehat{\mathfrak{J}}})$, namely
the \textit{conformal} Lie algebra $\mathfrak{g}=\mathfrak{conf}(\widehat{%
\mathfrak{J}})$ of $\widehat{\mathfrak{J}}$ itself (see \textit{e.g.} \cite%
{G-Lects} and \cite{SAM-Lectures} for a review, and a list of Refs.).

Interestingly, $\Delta $ also occurs in the duality-invariant expression of
the cosmological constant of some $AdS_{4}$ vacua (and of the corresponding
central charge of the dual CFT's) of general $\mathcal{N}=2$ \textit{gauged}
supergravities underlying flux compactifications of type $II$ theories \cite%
{Cassani:2009na}.

The fact that $f_{(abcd)}\neq 0$ which allows for the existence of
(primitive) \textit{quartic} $\mathfrak{g}$-invariant structure $\Delta
(\phi )$ characterizes the pair $\left( \mathfrak{g=conf}(\widehat{\mathfrak{%
J}}),\mathbf{R}\right) $ as a (\textit{non-degenerate}) Lie algebra \textit{%
of type }$\mathfrak{e}_{7}$, defined axiomatically by the axioms ($a$)-($c$)
of \cite{brown}: $\mathbf{R}$ is a representation space of $\mathfrak{g}$
such that

\begin{description}
\item[($a$)] $\mathbf{R}$ possesses a \textit{non-degenerate},
skew-symmetric bilinear $\mathfrak{g}$-invariant form (\textit{cfr.} (\ref%
{sympl-def}) and (\ref{omega-def}));

\item[($b$)] $\mathbf{R}$ possesses a \textit{completely symmetric}, rank-$4$
$\mathfrak{g}$-invariant structure $f_{\left( abcd\right) }$ ( given by the
completely symmetric part of (\ref{lower-sympl})), which allows to define%
\begin{equation}
q\left( x,y,z,w\right) \equiv f_{\left( abcd\right)
}x^{a}y^{b}z^{c}w^{d}=2\Delta \left( x,y,z,w\right) ;  \label{q-def}
\end{equation}

\item[($c$)] by defining a ternary product $\mathbf{T}\left( x,y,z\right) $
on $\mathbf{R}$ as%
\begin{equation}
\left\langle \mathbf{T}\left( x,y,z\right) ,w\right\rangle \equiv q\left(
x,y,z,w\right) ,  \label{deff}
\end{equation}%
then one has%
\begin{equation}
3\left\langle \mathbf{T}\left( x,x,y\right) ,\mathbf{T}\left( y,y,y\right)
\right\rangle =\left\langle x,y\right\rangle q\left( x,y,y,y\right) .
\end{equation}%
Note that, from (\ref{q-def}) and (\ref{deff}), $\mathbf{T}\left(
x,y,z\right) $ is the the completely symmetric part of the triple product $T$
(\ref{T-def}) on $\KK\sim \mathbf{R}$.
\end{description}

Recently, the role of Lie algebras of type $\mathfrak{e}_{7}$ was
investigated in supergravity in some detail (see Sec. \ref{FGT-sugra}). In\
Sec. \ref{FGT-SC} Brown's definition of Lie algebras \textit{of type} $%
\mathfrak{e}_{7}$ \cite{brown} will be discussed in relation to \textit{FTS}
and \textit{Freudenthal gauge theory}.

From the \textit{FTS} axioms discussed in Subsecs. \ref{Subsec-FTS-def} and %
\ref{Subsec-f_abcd}, one can show that $\Delta (\phi )$ is \textit{invariant}
under the following transformation:%
\begin{equation}
\mathcal{F}:\left\{
\begin{array}{l}
\KK\rightarrow \KK; \\
\\
\phi \mapsto \mathcal{F}\left( \phi \right) \equiv \text{sgn}\left( \Delta
(\phi )\right) \frac{T(\phi )}{\sqrt{6\,\left\vert \lam\Delta (\phi
)\right\vert }}\equiv \tilde{\phi},%
\end{array}%
\right.  \label{dual-field}
\end{equation}%
namely that%
\begin{equation}
\Delta (\phi )=\Delta (\tilde{\phi}),  \label{Delta-inv}
\end{equation}%
The proof can be found in App. \ref{App-FD} (which generalizes the treatment
of \cite{duff}, in turn referring to \cite{brown}, to \textit{FTS} defined
by axioms ($i$)-($iv$); see also \cite{alessio}). In the physics literature,
the map $\mathcal{F}$ (\ref{dual-field}) has been called \textquotedblleft
\textit{Freudenthal Duality}\textquotedblright\ (or \textit{F-duality} for
short); it was first observed in \cite{duff} as a symmetry of the
Bekenstein-Hawking \cite{BH-1,BH-2} entropy-area formula for black holes,
and then further generalized\footnote{%
In the nomenclature introduced in \cite{alessio}, (\ref{dual-field}) (which
preserves the homogeneity in $\phi $) defines the \textit{non-polynomial
\textquotedblleft on-shell"} version of \textit{F-duality}; other possible
versions and generalizations are discussed therein.} in \cite{alessio}%
.\smallskip

In the rest of this Subsection, we list some brief remarks; further details
will be reported in a forthcoming paper \cite{MST-2}.

\begin{itemize}
\item[(\textbf{I})] \textbf{Anti-Involutivity.} The \textit{F-duality} $%
\mathcal{F}$ (\ref{dual-field}) is an \textit{anti-involution} in $\KK$ \cite%
{brown,duff,alessio}:%
\begin{equation}
\begin{array}{c}
\mathcal{F}\circ \mathcal{F}=-Id; \\
\\
\tilde{\tilde{\phi}}=-\phi .%
\end{array}
\label{anti-involutivity}
\end{equation}%
This holds whenever $\phi $ is an element in $\MM_{\mathfrak{J}}^{c}$, which
is the complement in $\KK$ of the submanifold (recall (\ref{MM-2}))%
\begin{equation}
\left. \MM_{\mathfrak{J}}\right\vert _{I=J}\equiv \{\phi \in \KK\,|\,\LL%
_{\phi \phi }\phi \equiv T\left( \phi \right) =0\}\subset \KK.  \label{MM-J}
\end{equation}%
In addition to this, for $\lam\neq 0$ and for any $\phi \in \KK$, the
\textit{F-duality} map and its image $\tilde{\phi}$ (namely, the \textit{%
\textquotedblleft F-dual"} scalar field) are defined \textit{iff} $\Delta
(\phi )\neq 0$. Whenever $\mbox{Inv}(\MM_{\mathfrak{J}})$ is non-empty and
thus its corresponding action determines a \textit{stratification} of the
symplectic vector space $\KK\left( \mathfrak{J}\right) \sim \mathbf{R}\left( %
\mbox{Inv}(\MM_{\mathfrak{J}})\right) $ (\textit{cfr.} (\ref{FTS-R-gen})),
this can also be equivalently stated as the requirement that $\phi $ belongs
to the rank-$4$ orbit of $\KK$ under the action of $\mbox{Inv}(\MM_{%
\mathfrak{J}})$ itself.

\item[(\textbf{II})] $\mathbb{Z}_{4}$\textbf{-Grading.} The \textit{%
anti-involutivity} (\ref{anti-involutivity}) of\ $\mathcal{F}$ yields a $%
\mathbb{Z}_{4}$-grading of the symplectic vector space $\KK$. This
interesting property will be investigated in \cite{MST-2}.

\item[(\textbf{III})] \textbf{\textit{F-Duality} is not an \textit{FTS}
Derivation.} The non-linear map over $\KK$ provided by \textit{F-duality} (%
\ref{dual-field}) is \textit{not} a \textit{derivation} with respect to the
triple product (\ref{T-def}) over $\KK$. Thus, such a mathematical structure
cannot be consistently used to define an infinitesimal transformation. This
means that the invariance (\ref{Delta-inv}) is rather a \textit{global}
symmetry (\textit{\textquotedblleft duality"}) of $\KK$, and thus a \textit{%
global} (\textit{off-shell}) symmetry of the corresponding gauge theory; see
next Sections.
\end{itemize}

\section{\label{Sec-Gauge}Freudenthal Gauge Theory (FGT)}

In the present Section, we will introduce the gauge theory based on the
\textit{FTS} discussed in Sec. \ref{Sec-FTS}. As anticipated, this theory,
whose consistent (bosonic) Lagrangian density is proposed in Subsec. \ref%
{Subsec-Minimal}, will be named \textit{\textquotedblleft Freudenthal Gauge
Theory"} (\textit{FGT}).

As it will become clear, our construction resembles very much the one of BLG
theory \cite{BLG,gustavsson}. However, we present here a detailed analysis,
also in order to make several remarks addressing the differences between
\textit{FGT} (and thus \textit{FTS}) and the triple systems-related gauge
theories, especially in $D=3$ (see the discussion in Sec. \ref{FGT-SC}).

\subsection{\label{Subsec-Global}From \textit{Global} Symmetry...}

We consider a real scalar field $\phi (x)$ valued in a \textit{FTS} $\KK$
over $\mathbb{R}$, and we aim at constructing a Lagrangian density
functional $\mathbf{L}\left[ \phi (x)\right] $ with the desired symmetry.

Clearly, $\mathbf{L}\left[ \phi (x)\right] $ must be a $\KK$-scalar, and
thus all its terms must be of the form%
\begin{eqnarray}
\mathbf{L}\left[ \phi (x)\right] &\sim &\a(\phi )\,\la f(\phi ),g(\phi )\ra,
\label{L-gen-form} \\
\a &:&\left\{
\begin{array}{l}
\KK\rightarrow \mathbb{R}; \\
\phi \left( x\right) \mapsto \a\left( \phi (x)\right) ;%
\end{array}%
\right.  \label{alpha-def} \\
f,~g &:&\left\{
\begin{array}{l}
\KK\rightarrow \KK; \\
\phi \left( x\right) \mapsto f\left( \phi (x)\right) ;~~\phi \left( x\right)
\mapsto g\left( \phi (x)\right) .%
\end{array}%
\right.  \label{f-g-def}
\end{eqnarray}%
At each point $x$ in space-time, $f\left( \phi (x)\right) $ and $g\left(
\phi (x)\right) $ are elements of the subalgebra $\KK_{\phi (x)}\subset \KK$
generated by the element $\phi (x)\in \KK$. More precisely, elements of $\KK%
_{\phi (x)}$ are homogeneous polynomials of odd degree in $\phi (x)$, with
the multiplication defined by the \textit{non-associative} (\textit{cfr.}
axiom ($iii$)) triple product $T$ (\ref{T-def}) over $\KK$.

The \textit{FTS} axiom ($iii$) (or equivalently ($iii^{\prime }$)), along
with the definition (\ref{LL-def}), allow for a consistent definition of an
\textit{infinitesimal} transformation $\LL_{\Lam}\in \mathfrak{sp}(\KK)$
(recall (\ref{LL-sympl-valued})), such that%
\begin{equation}
\left[ f\left( (Id+\LL_{\Lam})\phi (x)\right) -f\left( \phi (x)\right) %
\right] _{\text{linear\ order}}=\LL_{\Lam}f\left( \phi (x)\right) ,
\label{inf-global}
\end{equation}%
where the parameters of the transformation are denoted by%
\begin{equation}
\Lam\in \KK\otimes _{s}\KK.  \label{Lam-def}
\end{equation}%
Note that only elements in the \textit{symmetric} tensor product $\KK\otimes
_{s}\KK$ can generate a transformation $\LL_{\Lam}$, because the
antisymmetric part $\KK\otimes _{a}\KK$ is projected out by the symmetry
property under the exchange of the first two entries of the triple product $%
T $ (\textit{cfr.} axiom ($i$)).

Crucially, axiom ($iv$) (or equivalently ($iv^{\prime }$)) states that for
any $f(\phi )$, $g(\phi )\in \KK$, the symplectic product $\la f(\phi
),g(\phi )\ra$ (defined in (\ref{sympl-def}) and in (\ref{omega-def})) is
\textit{invariant} under $\LL_{\Lam}$:
\begin{equation}
\LL_{\Lam}\la f(\phi ),g(\phi )\ra=\la\LL_{\Lam}f(\phi ),g(\phi )\ra+\la %
f(\phi ),\LL_{\Lam}g(\phi )\ra=0.  \label{inv-sympl-product}
\end{equation}%
By the same argument, all $\KK$-scalar real functions $\a(\phi )$ (\ref%
{alpha-def}) are necessarily of this form, namely%
\begin{equation}
\a(\phi )\sim \la h(\phi ),l(\phi )\ra
\end{equation}%
for some functions $h(\phi )$ and $l(\phi )$ of the same kind as $f(\phi )$
and $g(\phi )$ defined in (\ref{f-g-def}).\medskip

Thus, one can conclude that any Lagrangian density functional $\mathbf{L}$
of the form (\ref{L-gen-form}) is \textit{invariant}\footnote{%
Note that no mentioning of invariance under (\textit{global}; \textit{cfr.}
point (\textbf{IV}) of Subsec. \ref{Subsec-FD-def}) \textit{Freudenthal
duality} $\mathcal{F}$ (\ref{dual-field}) (which will be a crucial
ingredient of FGT; see Subsec. \ref{Subsec-Minimal}) has been made so far;
indeed, it is immediate to check that the Lagrangian density functional $%
\mathbf{L}$ (\ref{L-gen-form}) is\textit{\ not} \textit{invariant} under $%
\mathcal{F}$ (\ref{dual-field}).} under the infinitesimal transformation (%
\ref{inf-global}). In other words, by the four axioms ($i$)-($iv$) of
\textit{FTS}, any Lagrangian $\mathbf{L}$ of the form (\ref{L-gen-form}) is
guaranteed to be \textit{invariant} under the \textit{global} symmetry
generated by $\LL_{\Lam}$ (\ref{inf-global}).

It should also be remarked here that the definitions (\ref{Delta-def}) and (%
\ref{dual-field}) imply that the \textit{F-dual} field $\tilde{\phi}(x)$ is
also an element of $\KK_{\phi (x)}$. Therefore, $\tilde{\phi}(x)$ transforms
in the very same way as $\phi (x)$ under the global symmetry $\LL_{\Lam}$ (%
\ref{inf-global}).

As already pointed out above, the invariance (\ref{inv-sympl-product}) of
the symplectic product $\la\cdot ,\cdot \ra$ (\ref{sympl-def}) in $\KK$
under the action of the infinitesimal transformation $\LL_{\Lam}$ implies
that the latter is not simply an element in $\mathfrak{gl}(\KK)$, but rather
it generally belongs to the Lie algebra $\mathfrak{g}$ (\ref{GZ-g})-(\ref%
{LL-sympl-valued}).


\subsection{\label{Subsec-Gauge}...to \textit{Gauge} Symmetry}

We will now proceed to gauge the \textit{global} symmetry introduced in
Subsec. \ref{Subsec-Global}, by promoting the infinitesimal generator $\Lam$
(\ref{Lam-def}) to be a function $\Lam(x)$ over space-time. Correspondingly,
this will identify $\mathfrak{g}$ (\ref{GZ-g})-(\ref{LL-sympl-valued}) as
the \textit{gauge} algebra.

As done in Subsec. \ref{Subsec-FTS-def}, by adopting a basis $\{e_{a}\}$ for
$\KK$, one can generally write down the gauge transformation of a $\KK$%
-valued scalar field $\phi (x)=\phi ^{a}(x)e_{a}$ in the following form
(recall (\ref{f-def-2})):%
\begin{equation}
\LL_{\Lam}\phi (x)=\Lam^{ab}(x)\LL_{e_{a}e_{b}}\phi (x)={f_{abc}}^{d}\Lam%
^{ab}(x)\phi ^{c}(x)e_{d},  \label{L-gen-form-2}
\end{equation}%
%
%
%
%
%
%
%
%
%
%
%
%
%
%
%
%
%
%
%
%
%
%
%
%
%
%
%
%
%
%
%
%
%
%
%
%
%
%
%
%
%
%
where $\Lam^{ab}(x)$ denotes the rank-$2$ tensor generating the gauge
transformation itself. Note that axiom ($i$) of \textit{FTS} implies that
such a tensor is \textit{symmetric} (\textit{cfr.} (\ref{symm-L-call})):%
\begin{equation}
\LL_{e_{a}e_{b}}=\LL_{e_{b}e_{a}}\Leftrightarrow \Lam^{ab}(x)=\Lam^{ba}(x),
\label{Lam-symm}
\end{equation}%
which is consistent with (\ref{Lam-def}). When $\Lam^{ab}$ is \textit{%
constant} over space-time, one consistently re-obtains the \textit{global}
symmetry considered in Subsec. \ref{Subsec-Global}.

By recalling (\ref{LL-sympl-valued}), one can define the linear operator $%
\hat{\Lam}\in \mathfrak{g}$ as\footnote{%
In the following treatment, we will often drop the explicit $x$-dependence
in order to simplify the notation, whenever confusion is unlikely to occur.}%
\begin{equation}
\widehat{\Lambda }_{b}^{\phantom{b}a}\equiv {f_{cdb}}^{a}\Lam^{cd},
\label{Lam-hat-def}
\end{equation}%
such that the gauge symmetry transformation (\ref{L-gen-form-2}) of a field $%
\phi (x)$ is nothing but a matrix multiplication by the linear operator $%
\hat{\Lam}$:%
\begin{equation}
\LL_{\Lam}\phi ^{a}=\widehat{\Lambda }_{b}^{\phantom{b}a}\phi ^{b}.
\label{L-gen-form-2-recast}
\end{equation}%
As discussed at the end of Subsec. \ref{Subsec-Global}, the gauge
transformation of the \textit{F-dual} field $\tilde{\phi}(x)$ (\ref%
{dual-field}) is by construction the following one:%
\begin{equation}
\LL_{\Lam}\tilde{\phi}^{a}=\widehat{\Lambda }_{b}^{\phantom{b}a}\tilde{\phi}%
^{b}.  \label{L-gen-form-2-recast-dual}
\end{equation}

Next, we introduce a \textit{gauge field}%
\begin{equation}
A_{\m}(x)\equiv A_{\m}^{ab}(x)\,e_{a}\otimes _{s}e_{b},  \label{gauge-field}
\end{equation}%
which is a $1$-form valued in\footnote{%
Note that the \textit{symmetric} nature of the tensor product in (\ref%
{gauge-field}) does not imply any loss of generality, due to the axiom ($i$)
of \textit{FTS} (yielding $f_{cdb}^{\phantom{cdb}a}=f_{\left( cd\right)
b}^{~~~~~a}$).} $\KK\otimes _{s}\KK$. Correspondingly, a $\mathfrak{g}$%
-valued \textit{gauge covariant derivative} $D_{\m}$ acting on the scalar
field $\phi ^{a}(x)$ can be defined as:%
\begin{equation}
D_{\m}\phi ^{a}(x)\equiv \del_{\m}\phi ^{a}(x)-(\widehat{A}_{\mu })_{b}^{%
\phantom{b}a}(x)\phi ^{b}(x),  \label{gauge-cov-D}
\end{equation}%
where%
\begin{equation}
(\widehat{A}_{\mu })_{b}^{\phantom{b}a}(x)\equiv f_{cdb}^{\phantom{cdb}a}A_{%
\m}^{cd}(x)  \label{gauge-field-hat}
\end{equation}%
is the corresponding $1$-form linear operator in $\mathfrak{g}$.

It is worth remarking that both definitions (\ref{Lam-hat-def}) and (\ref%
{gauge-field-hat}) can respectively be regarded as images of the rank-$2$
symmetric tensor $\Lam^{ab}\left( x\right) $ (\ref{Lam-def}) of
infinitesimal gauge parameters and of the corresponding rank-$2$ symmetric
tensor $A_{\m}^{ab}(x)$ (\ref{gauge-field}) of $1$-form gauge potentials,
under a\textit{\ }map (dubbed \textit{\textquotedblleft hat"} map), defined
through the \textit{FTS structure constants} $f_{abc}^{\phantom{cdb}d}$ (\ref%
{f-def-2}) as follows:%
\begin{equation}
~\widehat{\cdot }:\left\{
\begin{array}{l}
\KK\otimes _{s}\KK\rightarrow \mathfrak{g}; \\
\\
\Psi ^{ab}(x)\,e_{a}\otimes _{s}e_{b}\mapsto f_{cdb}^{\phantom{cdb}a}\Psi
^{cd}(x)\equiv \widehat{\Psi }_{b}^{~a}.%
\end{array}%
\right.  \label{hat-map}
\end{equation}%
The \textit{\textquotedblleft hat"} map (\ref{hat-map}) allows one to
implement (generally $\mathfrak{g}$-valued) infinitesimal gauge
transformation $\LL_{\Lam}$ defined via the \textit{FTS} triple product in
terms of standard matrix multiplication (in $\mathfrak{gl}(\KK)$). As such,
this map provides an explicit matrix realization of the \textit{gauge Lie
algebra} $\mathfrak{g}$ of the FGT, by means of an embedding (\textit{local}
in space-time) analogous to the \textit{local} embedding $\KK_{\phi
(x)}\subset \KK$ mentioned below (\ref{f-g-def}).

Then, the requirement of $D_{\m}\phi (x)$ to transform under the gauge
symmetry $\LL_{\Lam}$ in the same way as $\phi (x)$, \textit{i.e.}%
\begin{equation}
\LL_{\Lam}\,\left( D_{\m}\phi ^{a}(x)\right) =(\LL_{\Lam}D_{\m})\phi
^{a}(x)+D_{\m}(\LL_{\Lam}\phi )^{a}(x)\equiv \widehat{\Lam}_{b}^{\phantom{b}%
a}(x)(D_{\m}\phi )^{b}(x)  \label{req-1}
\end{equation}%
consistently fixes the gauge transformation $\widehat{A}_{\m}(x)$ as follows:%
\begin{equation}
\LL_{\Lam}\widehat{A}_{\m}(x)=\del_{\m}\widehat{\Lam}(x)-\big[\widehat{A}_{\m%
}(x),\widehat{\Lam}(x)\big]\equiv D_{\m}\widehat{\Lam}(x),
\label{gauge-A-transf}
\end{equation}%
namely $\widehat{A}_{\m}(x)$ transforms as a $\mathfrak{g}$-valued $1$-form.

To proceed further, we introduce the gauge field strength $2$-form%
\begin{equation}
\widehat{F}_{\m\n}\equiv -\lbrack D_{\m},D_{\n}]=\del_{\m}\widehat{A}_{\n}-%
\del_{\n}\widehat{A}_{\m}-[\widehat{A}_{\m},\widehat{A}_{\n}]\in \mathfrak{g}%
,  \label{F-hat-def}
\end{equation}%
whose infinitesimal gauge transformation can consistently be computed to be%
\begin{equation}
\LL_{\Lam}\widehat{F}_{\m\n}=[\widehat{F}_{\m\n},\widehat{\Lam}].
\label{F-hat-gauge-transf}
\end{equation}%
The matrix embedding of $\LL_{\Lam}$ into $\mathfrak{g}$ provided by the
\textit{\textquotedblleft hat"} map (\ref{hat-map}) also ensures that the
\textquotedblleft trace\textquotedblright\ of the field strength $\widehat{F}%
_{\m\n}(x)$ (\ref{F-hat-def}) is $\mathfrak{g}$-\textit{gauge invariant}; in
the next Subsection, this fact will be used to work out a bosonic Lagrangian
for FGT.

\subsection{\label{Subsec-Minimal}The Lagrangian}

We are now going to propose a consistent bosonic Lagrangian for the FGT.

By recalling definitions (\ref{Delta-def}) and (\ref{dual-field}) and
considering the lowest possible order in the scalar field $\phi (x)$, one
can introduce the following (generally non-polynomial) term
\begin{equation}
\la\phi ,\tilde{\phi}\ra=\text{sgn}\left( \Delta (\phi )\right) \frac{\la%
\phi ,T(\phi )\ra}{\sqrt{6\left\vert \,\lam\Delta (\phi )\right\vert }}=-%
\sqrt{\frac{2}{3\left\vert \lam\right\vert }}\sqrt{\left\vert \Delta (\phi
)\right\vert },  \label{lowest-order-phi}
\end{equation}%
which is homogeneous of degree $2$ in $\phi (x)$. As discussed in Subsec. %
\ref{Subsec-Gauge}, the gauge covariant derivatives of both $\phi (x)$ and
its \textit{F-dual} field $\tilde{\phi}(x)$ transform as vectors under the
gauge transformation $\LL_{\Lam}$; therefore, a consistent kinetic term for
scalar fields reads%
\begin{equation}
-\frac{1}{2}\,\la D_{\m}\phi ,D^{\m}\tilde{\phi}\ra,  \label{kinetic-scalar}
\end{equation}%
whose \textit{gauge invariance} is guaranteed by the \textit{FTS} axioms ($i$%
)-($iv$), (\ref{inv-sympl-product}), and by the very treatment of Subsec. %
\ref{Subsec-Gauge}.\smallskip

From axiom ($iv$) (or equivalently (\ref{inv-sympl-product})) and (\ref%
{lowest-order-phi}), it follows that for any sufficiently smooth function $V:%
\mathbb{R\rightarrow R}$, then\footnote{%
Actually, by recalling definitions (\ref{alpha-def}) and (\ref{f-g-def}),
one could have chosen\linebreak\ $V\big(\a(\phi )\,\la f(\phi ),g(\phi )\ra%
\big)$ as the most general \textit{gauge invariant} potential term. However,
the invariance also under \textit{F-duality} $\mathcal{F}$ (\ref{dual-field}%
), as we do impose in FGT (see further below), further restricts the choice
to $V\big(\Delta (\phi )\big)$, as given by (\ref{potential-gauge-inv}%
).\smallskip}%
\begin{equation}
V\big(\Delta (\phi )\big)  \label{potential-gauge-inv}
\end{equation}%
is a \textit{gauge invariant} real function of $\phi $:%
\begin{equation}
\LL_{\Lam}\left( V\big(\Delta (\phi )\big)\right) =0,
\label{potential-gauge-inv-2}
\end{equation}%
which therefore can be taken as a \textit{gauge invariant} potential in the
bosonic FGT action.

By exploiting the matrix embedding of $\mathfrak{g}$-valued \textit{%
Freudenthal gauge transformations} $\LL_{\Lam}$ (realized by the \textit{%
\textquotedblleft hat"} map (\ref{hat-map})), one can construct a Maxwell
term for the \textit{gauge invariant} kinetic term for the gauge field $\hat{%
A}_{\m}(x)$.

By introducing the Minkowski metric $\eta _{\mu \nu }=\eta ^{\mu \nu }$ and
a function $\mathcal{N}\left( \Delta (\phi )\right) $ coupling vector and
scalar fields, for $D\geqslant 4$ the following kinetic Maxwell term can be
constructed:%
\begin{eqnarray}
\frac{1}{4}\mathcal{N}\left( \Delta (\phi )\right) \text{Tr}\left( \widehat{F%
}^{2}\right) &\equiv &\frac{1}{4}\mathcal{N}\left( \Delta (\phi )\right) %
\big(\widehat{F}_{\mu \nu }\big)_{a}^{\phantom{a}b}\big(\widehat{F}^{\mu \nu
}\big)_{b}^{\phantom{b}a}  \notag \\
&=&\frac{1}{4}\mathcal{N}\left( \Delta (\phi )\right) \eta ^{\mu \lambda
}\eta ^{\nu \rho }\,f_{cda}^{\phantom{cda}b}f_{efb}^{\phantom{efb}a}\,F_{\mu
\nu }^{cd}\,F_{\lambda \rho }^{ef}  \notag \\
&=&-\frac{1}{4}\mathcal{N}\left( \Delta (\phi )\right) \eta ^{\mu \lambda
}\eta ^{\nu \rho }\,f_{cdag}f_{efbh}\omega ^{ah}\omega ^{gb}\,F_{\mu \nu
}^{cd}\,F_{\lambda \rho }^{ef}.  \label{kinetic-vector}
\end{eqnarray}%
The \textit{gauge invariance} of (\ref{kinetic-vector}) results from the
simple computation
\begin{eqnarray}
\LL_{\Lam}\left( \frac{1}{4}\mathcal{N}\left( \Delta (\phi )\right) \text{Tr}%
\left( \widehat{F}^{2}\right) \right) &=&\frac{1}{4}\LL_{\Lam}\left(
\mathcal{N}\left( \Delta (\phi )\right) \right) \text{Tr}\left( \widehat{F}%
^{2}\right) +\frac{1}{4}\mathcal{N}\left( \Delta (\phi )\right) \LL_{\Lam%
}\left( \text{Tr}\left( \widehat{F}^{2}\right) \right) \\
&=&\frac{1}{2}\mathcal{N}\left( \Delta (\phi )\right) \text{Tr}\Big(\lbrack
\widehat{F},\widehat{\Lam}]\widehat{F}\Big)=0,  \label{kinetic-gauge-inv}
\end{eqnarray}%
where (\ref{potential-gauge-inv-2}) has been used for the function $\mathcal{%
N}$, the field strength gauge transformation property (\ref%
{F-hat-gauge-transf}) has been recalled, and the cyclicity of the trace has
been exploited.\medskip

Thus, by merging (\ref{kinetic-scalar}), (\ref{potential-gauge-inv}) and (%
\ref{kinetic-vector}), the following (bosonic) Lagrangian for the \textit{%
\textquotedblleft Freudenthal gauge theory" }(FGT) can be written down:%
\begin{equation}
\mathbf{L}\left[ \phi (x),F_{\mu \nu }\left( x\right) \right] _{D\geqslant
4}=-\frac{1}{2}\,\la D_{\m}\phi ,D^{\m}\tilde{\phi}\ra+\frac{1}{4}\mathcal{N}%
\left( \Delta (\phi )\right) \text{Tr}\left( \widehat{F}^{2}\right) -V\big(%
\Delta (\phi )\big),  \label{FGT-bos-action}
\end{equation}%
whose simplest (\textit{\textquotedblleft minimal"}) version corresponds to
setting $V\big(\Delta (\phi )\big)=\Delta (\phi )$ (\textit{quartic} scalar
potential) and $\mathcal{N}\left( \Delta (\phi )\right) =1$:
\begin{equation}
\mathbf{L}_{\text{minimal}}\left[ \phi (x),F_{\mu \nu }\left( x\right) %
\right] _{D\geqslant 4}=-\frac{1}{2}\,\la D_{\m}\phi ,D^{\m}\tilde{\phi}\ra+%
\frac{1}{4}\text{Tr}\left( \widehat{F}^{2}\right) -\Delta (\phi ).
\label{FGT-bos-action-min}
\end{equation}

Remarkably, the FGT Lagrangian density functional $\mathbf{L}\left[ \phi
(x),F_{\mu \nu }\left( x\right) \right] _{D\geqslant 4}$ (\ref%
{FGT-bos-action}) is not only \textit{invariant} under the \textit{off-shell}
gauge Lie algebra $\mathfrak{g}$ introduced in Subsecs. \ref{Subsec-Global}-(%
\ref{Subsec-Gauge}), but also under the \textit{F-duality} $\mathcal{F}$ (%
\ref{dual-field}), which acts as a \textit{global} (\textit{off-shell})
symmetry\footnote{%
From point (\textbf{IV}) of Subsec. \ref{Subsec-FD-def}), the \textit{%
Freudenthal duality} $\mathcal{F}$ (\ref{dual-field}) is not a \textit{%
derivation} with respect to the \textit{FTS} triple product (\ref{T-def})
over $\KK$, and thus with respect to the \textit{FTS}-based\textit{\ gauge}
transformation introduced above.}. In order to check this, one should simply
recall (\ref{Delta-inv}), as well as the \textit{anti-involutivity} (\ref%
{anti-involutivity}) of $\mathcal{F}$ (\ref{dual-field}) itself and the
anti-symmetry of the symplectic product used to construct the scalar kinetic
term (\ref{kinetic-scalar}). In particular, the $\mathcal{F}$-invariance of
the latter reads (recall point (\textbf{IV}) of Subsec. \ref{Subsec-FD-def}):%
\begin{eqnarray}
\mathcal{F}\,\Big(\eta ^{\m\n}\la D_{\mu }\phi ,D_{\n}\tilde{\phi}\ra\Big) %
&=&\eta ^{\m\n}\la D_{\mu }\tilde{\phi},D_{\n}(-\phi )\ra=\eta ^{\m\n}\la D_{%
\n}\phi ,D_{\mu }\tilde{\phi}\ra  \notag \\
&=&\eta ^{\m\n}\la D_{\mu }\phi ,D_{\n}\tilde{\phi}\ra,
\end{eqnarray}%
where in the second line one does not necessarily have to use the the
symmetry of the Minkowski space-time metric $\eta ^{\m\n}$, because, the
scalar kinetic term is symmetric under the exchange of its space-time
indices:%
\begin{equation}
\la D_{\mu }\phi ,D_{\nu }\tilde{\phi}\ra=\la D_{\nu }\phi ,D_{\mu }\tilde{%
\phi}\ra,  \label{symm-scalar-kin}
\end{equation}%
as shown in App. \ref{App-Symm-Scalar-Kinetic}.\medskip

It should be remarked here that in the above construction the dimension $D$
of space-time does not necessarily need to be specified. As mentioned, the ($%
\phi $-coupled) Maxwell kinetic vector term (\ref{kinetic-vector}) is well
defined in $D\geqslant 4$. Moreover, in $D=4$ a \textit{topological} (theta)
term can also be introduced, along with its vector-scalar coupling function $%
\mathcal{M}\left( \Delta (\phi )\right) $:%
\begin{equation}
\frac{1}{4}\mathcal{M}\left( \Delta (\phi )\right) \text{Tr}~\left( \widehat{%
F}\wedge \widehat{F}\right) ,  \label{theta-term}
\end{equation}%
and its \textit{gauge invariance} and $\mathcal{F}$\textit{-invariance} once
again follow from (\ref{potential-gauge-inv-2}), (\ref{F-hat-gauge-transf}),
(\ref{Delta-inv}) and the the cyclicity of the trace.

Thus, in $D=4$, the bosonic Lagrangian density (\ref{FGT-bos-action}) can be
completed as follows:%
\begin{eqnarray}
\mathbf{L}\left[ \phi (x),F_{\mu \nu }\left( x\right) \right] _{D=4} &=&-%
\frac{1}{2}\,\la D_{\m}\phi ,D^{\m}\tilde{\phi}\ra-V\big(\Delta (\phi )\big)
\notag \\
&&+\frac{1}{4}\mathcal{N}\left( \Delta (\phi )\right) \text{Tr}\left(
\widehat{F}^{2}\right) +\frac{1}{4}\mathcal{M}\left( \Delta (\phi )\right)
\text{Tr}~\left( \widehat{F}\wedge \widehat{F}\right) .  \label{FGT-D=4}
\end{eqnarray}

Even if in the above construction the dimension $D$ of space-time does not
necessarily need to be specified, it should be stressed that in $D\geqslant
4 $ the FGT is \textit{non-unitary} whenever the gauge Lie algebra $%
\mathfrak{g}$ is \textit{non-compact} (and thus with a Cartan-Killing metric
which is \textit{not} positive-definite). Indeed, we recall that in the
present investigation we consider the FTS to be defined on the ground field $%
\mathbb{R}$ (\textit{cfr.} Footnote 1); this constrains the pair $(\mathfrak{%
g},\mathbf{R})$ such that $\mathbf{R}$ is a real representation space of the
real algebra $\mathfrak{g}$. The latter, \textit{at least} in the examples
related to conformal symmetries of \textit{JTS} $\mathfrak{J}$ $=$ $\widehat{%
\mathfrak{J}}$ (treated in Sec. \ref{Gauge-Algebras-type-e7} and reported in
Table 1), is \textit{non-compact}.

On the other hand, in $D=3$ space-time dimensions this does not hold any
more, and the non-compactness of the (real) gauge Lie algebra $\mathfrak{g}$
is not inconsistent with unitarity of the theory. Indeed, $\mathbf{R}$ is
always assumed to possess a positive-definite inner product (for unitarity
of the corresponding gauge theory), but the gauge fields are not propagating
(and they are in $\mathbf{Adj}(\mathfrak{g})$), and therefore $\mathfrak{g}$
does not necessarily have to be endowed with a positive-definite product,
thus allowing for non-compact (real) forms of $\mathfrak{g}$ itself. As we
discuss in\ Sec. \ref{FGT-SC}, this is particularly relevant for the
connection between $D=3$ FGT and (the bosonic sector of) superconformal
Chern-Simons-matter gauge theories in $D=3$.

Moreover, in $D=3$ a Chern-Simons (CS) term for the gauge sector can be
considered, with the same form as in the BLG theory (\textit{cfr.} (45) of
\cite{BLG}):%
\begin{equation}
\frac{1}{2}\vep^{\m\n\lam}\Big(f_{abcd}\,A_{\m}^{ab}\,\del_{\n}A_{\lam}^{cd}+%
\frac{2}{3}f_{cda}^{\phantom{cda}g}f_{efgb}\,A_{\m}^{ab}A_{\n}^{cd}A_{\lam%
}^{ef}\Big),  \label{CS}
\end{equation}%
whose consistence in FGT follows from \textit{FTS} axioms ($i$) and ($iv$).
The $\mathcal{F}$\textit{-invariance} of the CS term (\ref{CS}) is trivial
(it does not depend on $\phi $ at all), while its \textit{gauge invariance}
can be easily proved by exploiting the symmetry property (\ref{symm-f}) of
\textit{FTS} structure constants $f_{abcd}$.

Thus, in $D=3$ one can propose the following bosonic FGT Lagrangian density:%
\begin{eqnarray}
\mathbf{L}\left[ \phi (x),F_{\mu \nu }\left( x\right) \right] _{D=3} &=&-%
\frac{1}{2}\,\la D_{\m}\phi ,D^{\m}\tilde{\phi}\ra-V\big(\Delta (\phi )\big)
\notag \\
&&+\frac{1}{2}\vep^{\m\n\lam}\Big(f_{abcd}\,A_{\m}^{ab}\,\del_{\n}A_{\lam%
}^{cd}+\frac{2}{3}f_{cda}^{\phantom{cda}g}f_{efgb}\,A_{\m}^{ab}A_{\n}^{cd}A_{%
\lam}^{ef}\Big).  \label{FGT-D=3}
\end{eqnarray}

\subsection{\label{Gauge-Algebras-type-e7}Gauge Algebras of Type $\mathfrak{e%
}_{7}$}

An interesting class of gauge algebras $\mathfrak{g}$ (\ref{GZ-g})-(\ref%
{LL-sympl-valued}) for the FGT can be obtained by considering symmetry
algebras of Jordan algebras $\widehat{\mathfrak{J}}$ themselves. Indeed, a
particular non-compact, real form of the decomposition (\ref{exc-Lie}) reads%
\begin{equation}
\mathfrak{qconf}(\widehat{\mathfrak{J}})=\mathfrak{conf}(\widehat{\mathfrak{J%
}})\oplus \mathfrak{sl}(2,\mathbb{R})\oplus \KK(\widehat{\mathfrak{J}}%
)\oplus \KK(\widehat{\mathfrak{J}}),  \label{exc-Lie-reinterpr}
\end{equation}%
where $\mathfrak{conf}(\widehat{\mathfrak{J}})$ and $\mathfrak{qconf}(%
\widehat{\mathfrak{J}})$ respectively denote the \textit{conformal} and
\textit{quasi-conformal}\footnote{%
The novel, non-linear geometric quasi-conformal realizations of groups were
first discovered by G\"{u}naydin, Koepsell and Nicolai in \cite{G8}, by
exploiting the underlying \textit{FTS}, and showing that they extend to the
complex forms and hence to different real forms of the corresponding groups.
In the subsequent papers \cite{G9} and \cite{GP}, the quasi-conformal
realizations of $D=3$ $U$-duality groups of Maxwell-Einstein supergravity
theories, respectively with with $8$ and at least $16$ supersymmetries, have
been determined. See \textit{e.g.} \cite{G-Lects}, for a review and a list
of Refs..} Lie algebras of rank-$3$ \textit{simple} Euclidean \textit{Jordan
algebras} $\widehat{\mathfrak{J}}$. Note that $\mathfrak{conf}(\widehat{%
\mathfrak{J}})$ is nothing but a particular non-compact, real form of $%
\mbox{Inv}(\MM_{\widehat{\mathfrak{J}}})$; this is also consistent with the
fact that $\mathfrak{conf}(\widehat{\mathfrak{J}})$ is nothing but the
\textit{automorphism} Lie algebra of $\KK(\widehat{\mathfrak{J}})$ itself:%
\begin{equation}
\mathfrak{conf}(\widehat{\mathfrak{J}})\sim \mathfrak{aut}\left( \KK(%
\widehat{\mathfrak{J}})\right) .
\end{equation}%
Analogously, also formul\ae\ (\ref{FTS-R})-(\ref{inters}) hold at the
suitable non-compact real level, by respectively replacing $\mbox{Inv}(\MM_{%
\widehat{\mathfrak{J}}})$ and $\mathfrak{sp}\left( \KK(\widehat{\mathfrak{J}}%
)\right) $ with $\mathfrak{conf}(\widehat{\mathfrak{J}})$ and\footnote{%
Note that $\mathfrak{sp}\left( f,\mathbb{R}\right) $ is the \textit{%
maximally non-compact} (\textit{split}) real form of $\mathfrak{sp}\left( \KK%
(\widehat{\mathfrak{J}})\right) $.} $\mathfrak{sp}\left( f,\mathbb{R}\right)
$. In particular, (\ref{inters}) can be recast as%
\begin{equation}
\mathfrak{conf}(\widehat{\mathfrak{J}})=\mathfrak{sp}\left( f,\mathbb{R}%
\right) \cap \mathfrak{qconf}(\widehat{\mathfrak{J}}).  \label{inters-2}
\end{equation}

The decompositions (\ref{exc-Lie}) and (\ref{exc-Lie-reinterpr}), as well as
the whole treatment above, also hold for rank-$3$ \textit{semi-simple}
Euclidean Jordan algebras of the type%
\begin{equation}
\widehat{\mathfrak{J}}=\mathbb{R\oplus }\mathbf{\Gamma }_{m,n},
\label{r=3-semi-simple}
\end{equation}%
where $\mathbf{\Gamma }_{m,n}$ is a rank-$2$ Jordan algebra with a quadratic
form of pseudo-Euclidean signature $(m,n)$, \textit{i.e.} the Clifford
algebra of $O(m,n)$ \cite{Jordan:1933vh}. However, in this case the
corresponding Lie algebra $\mathfrak{G}$ in (\ref{exc-Lie}) (or $\mathfrak{%
qconf}(\widehat{\mathfrak{J}})$ in (\ref{exc-Lie-reinterpr})) is a \textit{%
classical} Lie algebra, namely a (pseudo-)orthogonal algebra.

Table 1 lists the entries of (\ref{exc-Lie-reinterpr}) for rank-$3$
Euclidean Jordan algebras, also including the cases $\widehat{\mathfrak{J}}%
=J_{3}^{\mathfrak{A}_{s}}\equiv H_{3}(\mathfrak{A}_{s})$, where $\mathfrak{A}%
_{s}=\mathbb{C}_{s}$, $\mathbb{H}_{s}$, $\mathbb{O}_{s}$ are the \textit{%
split} version of $\mathbb{C}$, $\mathbb{H}$, $\mathbb{O}$, respectively
(see \textit{e.g.} \cite{G-Lects} for further elucidation and list of
Refs.). The role of $\KK(\widehat{\mathfrak{J}})$'s and their symmetries in
supergravity is discussed in the next Subsec. \ref{FGT-sugra}.

It is also worth recalling here that the Lie algebra $\mbox{Inv}(\MM_{%
\widehat{\mathfrak{J}}})$ (or equivalently $\mathfrak{conf}(\widehat{%
\mathfrak{J}})$) is \textit{\textquotedblleft of type }$\mathfrak{e}_{7}$%
\textit{"} \cite{brown}, as recalled in Sec. \ref{Subsec-FD-def}, and in the
mathematical literature its \textit{symplectic} (real) representation $%
\mathbf{R}$ is sometimes called \textit{minuscule} irrep. (see \textit{e.g.}
\cite{Manivel-1}).

\begin{table}[t]
\begin{center}
\begin{tabular}{|c||c|c|c|c|}
\hline
\rule[-1mm]{0mm}{6mm} $\widehat{\mathfrak{J}}$ & $\mathfrak{conf}(\widehat{%
\mathfrak{J}})$ & $\mathfrak{qconf}(\widehat{\mathfrak{J}})$ & $\mathbf{R}%
\left( \mathfrak{conf}(\widehat{\mathfrak{J}})\right) $ & $\mathcal{N}$ \\
\hline\hline
\rule[-1mm]{0mm}{6mm}$\mathbb{R}$ & $%
\begin{array}{ccc}
~ & \mathfrak{sl}(2,\mathbb{R}) & ~%
\end{array}%
$ & $%
\begin{array}{ccc}
~ & \mathfrak{g}_{2(2)} & ~%
\end{array}%
$ & $%
\begin{array}{ccc}
~ & \mathbf{4} & ~%
\end{array}%
$ & $%
\begin{array}{ccc}
~ & 2 & ~%
\end{array}%
$ \\ \hline
\rule[-1mm]{0mm}{6mm} \rule[-1mm]{0mm}{6mm}$\mathbb{R}\oplus $%
\rule[-1mm]{0mm}{6mm}$\mathbb{R}$ & $\mathfrak{sl}(2,\mathbb{R})\oplus
\mathfrak{sl}(2,\mathbb{R})$ & $\mathfrak{so}\left( 3,4\right) $ & $\left(
\mathbf{2},\mathbf{3}\right) $ & $2$ \\ \hline
\rule[-1mm]{0mm}{6mm} \rule[-1mm]{0mm}{6mm}$\mathbb{R}\oplus $%
\rule[-1mm]{0mm}{6mm}$\mathbb{R}\oplus $\rule[-1mm]{0mm}{6mm}$\mathbb{R}$ & $%
\mathfrak{sl}(2,\mathbb{R})\oplus \mathfrak{sl}(2,\mathbb{R})\oplus
\mathfrak{sl}(2,\mathbb{R})$ & $\mathfrak{so}(4,4)$ & $\left( \mathbf{2},%
\mathbf{2},\mathbf{2}\right) $ & $2$ \\ \hline
\rule[-1mm]{0mm}{6mm}$\mathbb{R\oplus }\mathbf{\Gamma }_{m,n}$ & $\mathfrak{%
sl}(2,\mathbb{R})\oplus \mathfrak{so}(m+1,n+1)$ & $\mathfrak{so}(m+2,n+2)$ &
$\left( \mathbf{2},\mathbf{m+n+2}\right) $ & \multicolumn{1}{|l|}{$%
\begin{array}{c}
2~\left( m=1\right) ~ \\
4~\left( m=5\right)%
\end{array}%
$} \\ \hline
$J_{3}^{\mathbb{R}}$ & $\mathfrak{sp}(6,\mathbb{R})$ & $\mathfrak{f}_{4(4)}$
& $\mathbf{14}^{\prime }$ & $2$ \\ \hline
$J_{3}^{\mathbb{C}}$ & $\mathfrak{su}(3,3)$ & $\mathfrak{e}_{6(2)}$ & $%
\mathbf{20}$ & $2$ \\ \hline
$J_{3}^{\mathbb{C}_{s}}$ & $\mathfrak{sl}(6,\mathbb{R})$ & $\mathfrak{e}%
_{6(6)}$ & $\mathbf{20}$ & $0$ \\ \hline
$M_{1,2}(\mathbb{O})$ & $\mathfrak{su}(1,5)$ & $\mathfrak{e}_{6(-14)}$ & $%
\mathbf{20}$ & $5$ \\ \hline
$J_{3}^{\mathbb{H}}$ & $\mathfrak{so}^{\ast }(12)$ & $\mathfrak{e}_{7(-5)}$
& $\mathbf{32}^{(\prime )}$ & $2,~6$ \\ \hline
$J_{3}^{\mathbb{H}_{s}}$ & $\mathfrak{so}(6,6)$ & $\mathfrak{e}_{7(7)}$ & $%
\mathbf{32}^{(\prime )}$ & $0$ \\ \hline
$J_{3}^{\mathbb{O}}$ & $\mathfrak{e}_{7(-25)}$ & $\mathfrak{e}_{8(-24)}$ & $%
\mathbf{56}$ & $2$ \\ \hline
$J_{3}^{\mathbb{O}_{s}}$ & $\mathfrak{e}_{7(7)}$ & $\mathfrak{e}_{8(8)}$ & $%
\mathbf{56}$ & $8$ \\ \hline
\end{tabular}%
\end{center}
\caption{\textit{Conformal} $\mathfrak{conf}(\protect\widehat{\mathfrak{J}})$
and \textit{quasi-conformal} $\mathfrak{qconf}(\protect\widehat{\mathfrak{J}}%
)$ Lie algebras associated to rank-$3$ Euclidean Jordan algebras. The
relevant symplectic irrep. $\mathbf{R}$ of $\mathfrak{conf}(\protect\widehat{%
\mathfrak{J}})$ is also reported. In particular, $\mathbf{14}^{\prime }$
denotes the rank-$3$ antisymmetric irrep. of $\mathfrak{sp}(6,\mathbb{R})$,
whereas $\mathbf{32}$ and $\mathbf{32}^{\prime }$ are the two chiral spinor
irreps. of $\mathfrak{so}^{\ast }\left( 12\right) .$ Note that $\mathfrak{%
conf}(J_{3}^{\mathfrak{A}_{s}})$ and $\mathfrak{qconf}(J_{3}^{\mathfrak{A}%
_{s}})$ are the \textit{maximally non-compact} (\textit{split}) real forms
of the corresponding compact Lie algebra. $M_{1,2}\left( \mathbb{O}\right) $
is the \textit{JTS} generated by $2\times 1$ vectors over $\mathbb{O}$
\protect\cite{Gunaydin:1983rk, Gunaydin:1983bi}. Note the Jordan algebraic
isomorphisms $\mathbf{\Gamma }_{1,1}\sim \mathbb{R\oplus R}$, and $\mathbf{%
\Gamma }_{1,0}\sim \mathbb{R}$. The number of spinor supercharges $\mathcal{N%
}$ of the corresponding supergravity theory in $D=4$ (\textit{cfr.} Subsec.
\protect\ref{FGT-sugra}) is also listed.}
\label{grouptheorytable}
\end{table}

\subsection{\label{FGT-sugra}FGT and Supergravity}

Summarizing, a class of gauge algebras (and representations) for FGT is
provided by the conformal Lie algebras $\mathfrak{conf}$ of (simple and
semi-simple) Euclidean, rank-$3$ algebras $\widehat{\mathfrak{J}}$, listed
in Table 1, along with their (real) symplectic representation $\mathbf{R}$.
The pair $\left( \mathfrak{conf}\left( \widehat{\mathfrak{J}}\right) ,%
\mathbf{R}\right) $ characterizes $\mathfrak{conf}\left( \widehat{\mathfrak{J%
}}\right) $ as a Lie algebra \textit{of type }$\mathfrak{e}_{7}$ \cite{brown}%
.

Interestingly, $\mathfrak{conf}\left( \widehat{\mathfrak{J}}\right) $ is the
$U$-duality\footnote{%
Here $U$-duality is referred to as the \textquotedblleft
continuous\textquotedblright\ symmetries of \cite{CJ-1}. Their discrete
versions are the $U$-duality non-perturbative string theory symmetries
introduced by Hull and Townsend \cite{HT-1}.} Lie algebra of $D=4$
Maxwell-Einstein supergravity theories (MESGT's) related to the FTS $\KK(%
\widehat{\mathfrak{J}})$ \cite{Gunaydin:1983rk, Gunaydin:1983bi} (see also
\textit{e.g.} \cite{G-Lects} and \cite{Borsten:2011ai, Borsten:2011nq} for
recent reviews, and list of Refs.).\medskip

Indeed, within such a class of theories, the decomposition (\ref%
{exc-Lie-reinterpr}) can be further interpreted as the Cartan decomposition
of the $\mathfrak{qconf}(\widehat{\mathfrak{J}})$ ($U$-duality algebra in $%
D=3$) with respect to $\mathfrak{conf}(\widehat{\mathfrak{J}})$ ($U$-duality
algebra in $D=4$). In particular, $\mathbf{R}\left( \mathfrak{conf}(\widehat{%
\mathfrak{J}})\right) $ listed in Table 1 is the representation in which the
$2$-form field strengths of the $D=4$ Abelian vector potentials sit, along
with their duals. As mentioned above, $\mathfrak{conf}(\widehat{\mathfrak{J}}%
)$ is nothing but $\mbox{Inv}(\MM_{\widehat{\mathfrak{J}}})$, possibly
specified as a suitable \textit{non-compact} real algebra\footnote{%
In fact, as a maximal subalgebra of $\mathfrak{qconf}(\widehat{\mathfrak{J}}%
) $, in this framework the Lie algebra $\mbox{Inv}(\MM_{\widehat{\mathfrak{J}%
}})$ can be compact (with commuting subalgebra $\mathfrak{su}(2)$) or
non-compact (with commuting subalgebra $\mathfrak{sl}(2,\mathbb{R})$),
depending on whether the Kaluza-Klein reduction from $D=4\rightarrow 3$ is
performed along a space-like or time-like direction, respectively; in turn,
this mathematically corresponds to perform a $c$\textit{-map} \cite%
{Cecotti:1988qn} or a $c^{\ast }$\textit{-map} (see \textit{e.g.} \cite{BGM}%
) on the $D=4$ (vector multiplets') scalar manifold.}.

\textit{At least} in $D=3,4,5,6$, the theories of this class all exhibit
(Abelian vector multiplets') scalar manifolds which are \textit{symmetric}
cosets\footnote{%
A particular case is given by $M_{1,2}\left( \mathbb{O}\right) $, which (%
\textit{cfr.} caption of Table 1) is a \textit{JTS} generated by $2\times 1$
vectors over $\mathbb{O}$ \cite{Gunaydin:1983rk, Gunaydin:1983bi}. It is
related to supergravity with $20$ local supersymmetries, which exists only
in $D=4$ ($\mathcal{N}=5$ \cite{N=5}) and in $D=3$ ($\mathcal{N}=10$; see
\textit{e.g.} \cite{Tollsten} and Refs. therein).}. In particular, the coset
Lie generators in $D=4$ and $D=3$ Lorentzian space-time dimensions are
respectively given by $\mathfrak{conf}(\widehat{\mathfrak{J}})$ and $%
\mathfrak{qconf}(\widehat{\mathfrak{J}})$ modded out by their maximal
compact subalgebra ($mcs$).

The number of spinor supercharges $\mathcal{N}$ of the $D=4$ supergravity
theory is reported in Table 1. In particular, the theories associated to $%
\widehat{\mathfrak{J}}=J_{3}^{\mathfrak{A}}\equiv H_{3}(\mathfrak{A})$ are
usually dubbed \textit{"magical"} MESGT's \cite{Gunaydin:1983rk,
Gunaydin:1983bi}, whereas the $\mathcal{N}=2$, $D=4$ theories corresponding
to $\widehat{\mathfrak{J}}=\mathbb{R}$, $\mathbb{R\oplus R}$ and $\mathbb{%
R\oplus R\oplus R}$ are the so-called $T^{3}$, $ST^{2}$ and $STU$ models
\cite{Duff:1995sm, Behrndt:1996hu}. It should also be remarked that $%
\widehat{\mathfrak{J}}=J_{3}^{\mathbb{H}}$ is related to both $\mathcal{N}=2$
and $\mathcal{N}=6$ theories, which in fact share the very same bosonic
sector \cite{Gunaydin:1983rk, Gunaydin:1983bi, Andrianopoli:1997pn,
Ferrara:2008ap, Roest:2009sn}.

As discussed in Subsec. \ref{Subsec-Lie-Algebras}, \textit{FTS}'s $\KK\left(
\widehat{\mathfrak{J}}\right) $ (with $\widehat{\mathfrak{J}}$ \textit{simple%
}) exhibit a close relationships with \textit{exceptional} Lie algebras, as
given by (\ref{exc-Lie}). As listed in Table 1, when considering suitable
non-compact, real forms, (\ref{exc-Lie}) enjoys the reinterpretation (\ref%
{exc-Lie-reinterpr}) : in other words, \textit{exceptional} Lie algebras
occur as \textit{quasi-conformal} Lie algebras of the corresponding \textit{%
simple} Jordan algebras $\widehat{\mathfrak{J}}$ \cite{freudenthal,yamaguti}%
. In this respect, it is worth adding that \textit{classical} (namely,
\textit{pseudo-othogonal}) Lie algebras also occur as \textit{quasi-conformal%
} Lie algebras of rank-$3$ \textit{semi-simple} Euclidean \textit{Jordan
algebras} of the type\footnote{%
The quasi-conformal realizations constructed in \cite{G8,G9,GP} correspond
to non-linear geometric constructions that leave invariant a generalized
lightcone with respect to a quartic distance function. As such, they are
different from the algebraic constructions of Lie algebras over triple
systems given in the mathematics literature (see \textit{e.g.} \cite%
{freudenthal,yamaguti}).} (\ref{r=3-semi-simple}) \cite{GP}.

These facts provide indication of possible links between \textit{FGT} and
Yang-Mills (\textit{exceptional}) gauge theories.\medskip

At bosonic level, differences and similarities between the \textit{FGT} and
the class of MESGT's under consideration can be observed by comparing
\textit{e.g.} the $D=3$ \textit{FGT} Lagrangian density (\ref{FGT-D=3}) with
the bosonic sector of the (ungauged) MESGT ($D=4$) Lagrangian density (%
\textit{cfr.} \textit{e.g.} the treatment in \cite{G10}, and Refs. therein)%
\begin{equation}
e^{-1}\mathcal{L}=-\frac{1}{2}R-g_{ij}\partial _{\mu }\phi ^{i}\partial
^{\mu }\phi ^{j}+\frac{1}{4}\text{Im}\left( \mathcal{N}_{\Lambda \Sigma
}\right) F_{\mu \nu }^{\Lambda }F^{\Sigma \mid \mu \nu }-\frac{e^{-1}}{8}%
\epsilon ^{\mu \nu \rho \sigma }\text{Re}\left( \mathcal{N}_{\Lambda \Sigma
}\right) F_{\mu \nu }^{\Lambda }F_{\rho \sigma }^{\Sigma }.
\end{equation}%
Besides the presence of the Einstein-Hilbert term, there are crucial
differences : in the \textit{FGT} the scalar fields $\phi $ fit into $%
\mathbf{R}$($\mathfrak{g}$) and the vectors arise from the gauging of the
\textit{FTS} triple product symmetry algebra $\mathfrak{g}$; as a
consequence, the derivatives acting on $\phi $ are covariantized, as
discussed in Secs. \ref{Subsec-Gauge} and \ref{Subsec-Minimal}. On the other
hand, in the corresponding ($D=4$) supergravity framework, the Abelian
two-form field strengths fit into $\mathbf{R}$($\mathfrak{g}=\mathfrak{conf}(%
\widehat{\mathfrak{J}})$), while the scalar fields are in a suitable
representation of the maximal compact subalgebra $mcs(\mathfrak{g})$.
Furthermore, as discussed above, in \textit{FGT} the gauge algebra $%
\mathfrak{g}=\mathfrak{conf}(\widehat{\mathfrak{J}})$ and the corresponding
global Freudenthal duality are \textit{off-shell} symmetries of the theory,
whereas in the MESGT's under consideration $\mathfrak{g}=\mathfrak{conf}(%
\widehat{\mathfrak{J}})$ is only an \textit{on--shell} symmetry\footnote{%
One can construct manifestly $U$-invariant Lagrangians, but at the price of
a non-manifest Lorentz-invariance \cite{NML} or of doubling the field
strengths' degrees of freedom (\textit{doubled formalism }\cite{DF}; for
recent advances in relation to Freudenthal duality, see \textit{e.g.} \cite%
{FDL}).}. It is also worth pointing out that on the gravity side
supersymmetry seems to be an accidental feature; indeed, we recall that for $%
\widehat{\mathfrak{J}}=J_{3}^{\mathbb{C}_{s}}$ and $J_{3}^{\mathbb{H}_{s}}$,
the corresponding theories of gravity coupled to Maxwell and scalar fields
are not supersymmetric; possible supersymmetrization of \textit{FGT} will be
discussed in Sec. \ref{Sec-Gen}.

It will be interesting to investigate these relations in future studies; see
also the discussion in Sec. \ref{FGT-SC}.

\section{\label{Sec-Gen}Generalization?}

In the previous Section, we have constructed a consistent Lagrangian for%
\textit{\ }the \textit{Freudenthal gauge theory} (FGT), based on the FTS\ $%
\KK\left( \mathfrak{J}\right) $, with $\KK$-valued scalar field $\phi (x)$,
admitting both (\textit{off-shell}) FTS \textit{gauge} symmetry and (\textit{%
off-shell}) \textit{global} \textit{Freudenthal-duality }symmetry\textit{\ }$%
\mathcal{F}$.

The most important kind of generalization would concern an FGT-type
Lagrangian involving some \textit{vector} fields \textit{and/or} \textit{%
spinor} fields, which is again invariant under both \textit{FTS} gauge and
\textit{Freudenthal duality} symmetries; indeed, this would be a necessary
condition for a supersymmetric (non-trivial) extension of FGT. Moreover,
such a generalization is of interest to the physicists, since it potentially
might define a \textit{sigma-model} type theory if the space-time considered
in this paper is regarded as the \textit{world-volume} of some extended
objects (for instance, $M2$-branes), and correspondingly the \textit{vector}
fields conceived as the image of the \textit{world-volume} in some target
space.

However, in Subsecs. \ref{Subsec-Couple}-\ref{Subsec-No--Go} we shall prove
that, within some minimal reasonable assumptions, such a generalization is
not possible.

\subsection{\label{Subsec-Couple}Coupling to a Vector Space}

Let us start the analysis by coupling a generic \textit{FTS} $\KK$ to a
generic vector space $\VV$, over which one can introduce suitable algebraic
structures and make it into an algebra; for instance, spinors can be
regarded as vectors with an anti-symmetric binary product that yields the
Fermi statistics. In this way, our discussion for the formal algebraic
system $\VV$ will cover the most generic space that couples to $\KK$.

Thus, we are considering an \textit{extended} vector space%
\begin{equation}
\NN\equiv \KK\otimes \VV,  \label{NN}
\end{equation}%
whose element, denoted by $\Phi $, is the tensor product of an element $\phi
\in \KK$ and an element $v\in \VV$, \textit{i.e.}%
\begin{equation}
\Phi \equiv \phi \otimes v\in \NN.  \label{Phi}
\end{equation}

In order to be able to construct a Lagrangian density functional $\mathbf{L}%
\left[ \Phi (x)\right] $ for the fields $\Phi (x)\in \NN$ obtained from
promoting an element $\Phi $ $\in \NN$ to a $\NN$-valued space-time field $%
\Phi (x)$, one starts by introducing a bilinear form (namely, the \textit{%
metric})%
\begin{equation}
\la\cdot ,\cdot \ra:\left\{
\begin{array}{l}
\NN\otimes \NN\rightarrow \mathbb{R}; \\
\\
\Phi _{I},\Phi _{J}\mapsto \la\Phi _{I},\Phi _{J}\ra,%
\end{array}%
\right.  \label{metric-def}
\end{equation}%
defined for any two $\Phi _{I,J}=\phi _{I,J}\otimes v_{I,J}$ in $\NN$.
\textit{Via} direct evaluation, (\ref{metric-def}) induces a \textit{metric}
on $\VV$ itself:%
\begin{equation}
\la\Phi _{I},\Phi _{J}\ra=\la\phi _{I}\otimes v_{I},\phi _{J}\otimes v_{J}\ra%
=\la\phi _{I},\phi _{J}\ra\times (v_{I},v_{J})_{_{\VV}},\quad \forall \,\Phi
_{I},\Phi _{J}\in \NN,
\end{equation}%
where \textquotedblleft $\times $\textquotedblright\ is here multiplication
by a scalar (real) factor, and
\begin{equation}
(\cdot ,\cdot )_{_{\VV}}:\left\{
\begin{array}{l}
\VV\otimes \VV\rightarrow \mathbb{R}; \\
\\
v_{I},v_{J}\mapsto (v_{I},v_{J})_{_{\VV}},%
\end{array}%
\right.  \label{ind-metric}
\end{equation}%
is the \textit{induced metric} over $\VV$. Note that the symmetry property
of $(\cdot ,\cdot )_{_{\VV}}$ (\ref{ind-metric}) is to be determined by the
required symmetry property of the metric $\la\cdot ,\cdot \ra$ (\ref%
{metric-def}) over $\NN$ (by also recalling the anti-symmetry of the
symplectic form (\ref{sympl-def}) over $\KK$).

Furthermore, in order to consistently define the \textit{Freudenthal duality
}$\mathcal{F}$ of this \textit{extended} theory, one needs to introduce a
triple product%
\begin{equation}
\mathcal{T}:\left\{
\begin{array}{l}
\NN\otimes \NN\otimes \NN\rightarrow \NN; \\
\\
\Phi _{I},\Phi _{J},\Phi _{K}\mapsto \mathcal{T}\left( \Phi _{I},\Phi
_{J},\Phi _{K}\right) \equiv \Phi _{I}\Phi _{J}\Phi _{K},%
\end{array}%
\right.  \label{T-call-def}
\end{equation}%
defined for any three elements $\Phi _{I},\Phi _{J},\Phi _{K}\in \NN$, which
would then induce a tri-linear \textit{triple product} on $\VV$ itself:%
\begin{equation}
\left[ \cdot ,\cdot ,\cdot \right] _{\VV}:\left\{
\begin{array}{l}
\VV\otimes \VV\otimes \VV\rightarrow \VV; \\
\\
v_{I},v_{J},v_{K}\mapsto \lbrack v_{I},v_{J},v_{K}]_{_{\VV}}.%
\end{array}%
\right.  \label{triple-VV}
\end{equation}

In order to proceed further, we make here a plausible conjecture that
\textit{Freudenthal duality }$\mathcal{F}$ can be defined only for algebraic
systems satisfying the axioms ($i$)-($iv$) of an \textit{FTS,} introduced in
Subsec. \ref{Subsec-FTS-def}. As a consequence, we require the metric (\ref%
{metric-def}) to be an \textit{anti-symmetric} bilinear form (and append
this as axiom $(o)$), thus obtaining the following five axioms for the
algebra $\NN$:

\begin{itemize}
\item[($o$)] $\la\Phi_I,\Phi_J\ra=-\la \Phi_J,\Phi_I\ra$

\item[($i$)] $\Phi_I\Phi_J\Phi_K=\Phi_J\Phi_I\Phi_K$

\item[($ii$)] $\Phi_I\Phi_J\Phi_K=\Phi_I\Phi_K\Phi_J+2\m\,\la \Phi_J,\Phi_K%
\ra\,\Phi_I +\m\,\la \Phi_I,\Phi_K\ra\,\Phi_J-\m\,\la \Phi_I,\Phi_J\ra%
\,\Phi_K$

\item[($iii$)] $\Phi_L\Phi_M(\Phi_I\Phi_J\Phi_K)=(\Phi_L\Phi_M\Phi_I)\Phi_J%
\Phi_K +\Phi_I(\Phi_L\Phi_M\Phi_J)\Phi_K+\Phi_I\Phi_J(\Phi_L\Phi_M\Phi_K)$

\item[($iv$)] $\la \Phi_L\Phi_M\Phi_I,\Phi_J\ra+\la\Phi_I, \Phi_L\Phi_M\Phi_J%
\ra=0$,
\end{itemize}

where $\m$ plays the role of the real parameter $\lam$ introduced above for
the \textit{FTS} $\KK$.

Then, by repeating for the algebra $\NN$ the very same construction
discussed in Sec. \ref{Sec-Gauge} for the \textit{FTS} $\KK$, one gets the
most general Lagrangian density functional $\mathbf{L}\left[ \Phi (x)\right]
$ invariant under the two desired symmetries, namely under both (\textit{%
off-shell}) FTS \textit{gauge} symmetry and (\textit{off-shell}) \textit{%
global} \textit{Freudenthal-duality }symmetry\textit{\ }$\mathcal{F}$.

\subsection{\label{Subsec-No--Go}A \textit{No-Go} Theorem}

However, this seemingly smooth construction of an \textit{extended} FGT
coupled to vector \textit{and/or} spinor fields suffers from some severe
constraints, which actually spoils the above generalization.

Indeed, axioms ($o$)-($iv$) of $\NN$ induce a set of corresponding axioms
for the metric (\ref{ind-metric}) and the triple product (\ref{triple-VV})
induced on $\VV$ (in addition to the ones already introduced for other
physical reasons, such as the ones yielded by the Bose \textit{and/or} Fermi
statistics for the fields $v_{I}\in $ $\VV$); the reader can find the full
set of such axioms for $\VV$ in App. \ref{App-Ind-Axioms}.

Among them, axiom ($B.\,iii$) induced from the \textit{derivation} property
of $\NN$ leads to a particularly strong constraint. In order to realize
this, let us restrict to a subalgebra%
\begin{equation}
\NN_{\phi }\equiv \KK_{\phi }\otimes \VV\subset \NN,  \label{sub-NN}
\end{equation}%
where $\KK_{\phi }$ is the subalgebra in $\KK$ generated by a single
generator $\phi \in \KK$ (see also Subsec. \ref{Subsec-Global}). Then, by
taking five elements of the form%
\begin{equation}
\Phi _{L,M,I,J,K}\equiv \phi \otimes v_{L,M,I,J,K}\in \NN_{\phi }
\end{equation}%
and inserting them into axiom ($B.\,iii$) of App. \ref{App-Ind-Axioms}, the
following simplified (weaker) condition on the algebraic structure of $\VV$
is achieved:%
\begin{equation}
\phi \phi \,T(\phi )\otimes \Big(\big[v_{L},v_{M},[v_{I},v_{J},v_{K}]_{_{\VV%
}}\big]_{_{\VV}}-\big[v_{I},v_{J},[v_{L},v_{M},v_{K}]_{_{\VV}}\big]_{_{\VV}}%
\Big)=0,  \label{weaker-cond}
\end{equation}%
where the simplification comes from the fact that over the subalgebra $\KK%
_{\phi }$, $\LL_{\phi \,T(\phi )}$ and $\LL_{T(\phi )\phi }$ act as \textit{%
annihilation operators}, whose proof can be found in App. \ref{App-FD}.

Moreover, we observe that, as holding for $\KK$ (\textit{cfr.} definition (%
\ref{LL-def})) for any two elements $v_{L}$, $v_{M}$ $\in $ $\VV$ one gets a
linear operator (generally $\mathfrak{gl}(\VV)$-valued, whenever it is
non-zero) $\LL_{v_{L}v_{M}}$, whose action is evaluated by the triple
product (\ref{triple-VV}) as:%
\begin{equation}
\LL_{v_{L}v_{M}}:\left\{
\begin{array}{l}
\VV\otimes \VV\rightarrow \VV; \\
\\
v_{L},v_{M}\mapsto \LL_{v_{L}v_{M}}\,v_{I}\equiv \lbrack
v_{L},v_{M},v_{I}]_{_{\VV}},%
\end{array}%
\right.  \label{LL-V-def}
\end{equation}%
Then, by using definition (\ref{LL-V-def}), the weaker form (\ref%
{weaker-cond}) of the axiom ($B.\,iii$) can be recast as a condition on the
matrix commutator in $\mathfrak{gl}(\VV)$:
\begin{equation}
\big[\LL_{v_{L}v_{M}},\LL_{v_{I}v_{J}}\big]=0,~\forall v_{I,J,L,M}\in \VV.
\label{weaker-cond-2}
\end{equation}

Under the assumption that the metric (\ref{metric-def}) in $\NN$ is \textit{%
non-degenerate} (which we understand throughout\footnote{%
For instance, if the metric (\ref{metric-def}) of the algebra $\NN$ is
\textit{degenerate}, it can be proved that a coupling to a Grassmannian
number degree of freedom is possible. However, since there is no fermion
bilinear for a one-dimensional spinor, this is a rather trivial theory, in
which the fermionic degree of freedom never plays any role, and it cannot
mix up with the bosonic degree of freedom. In such a theory, the structure
is essentially the same as the one pertaining to a single $\KK$-valued
(real) scalar field, and hence a consistent implementation of invariance
under (\textit{global}, \textit{off-shell}) \textit{Freudenthal duality} is
possible. We plan to investigate further this issue in future work.}), the
condition (\ref{weaker-cond-2}) can be satisfied in only two instances:

\begin{itemize}
\item[{[ $\mathbb{I}$ ]}] when dim$_{\left( \mathbb{R}\right) }\VV=1$,
\textit{i.e.}%
\begin{equation}
\NN=\KK\otimes \mathbb{R},  \label{possib-1}
\end{equation}%
which is the case of a single $\KK$-valued (real) scalar field discussed in
Secs. \ref{Sec-FTS}-\ref{Sec-Gauge};

\item[{[ $\mathbb{II}$ ]}] when the set%
\begin{equation}
\{\LL_{v_{I}v_{J}}\in \mathfrak{gl}(\VV)\,|\,v_{I},v_{J}\in \VV\}\subset
\mathfrak{gl}(\VV)  \label{set}
\end{equation}%
is a subset of the \textit{Cartan subalgebra} of $\mathfrak{gl}(\VV)$, namely%
\footnote{%
In general, instead of (\ref{possib-2}) one may propose%
\begin{equation*}
\LL_{v_{I}v_{J}}\,v_{K}=[v_{I},v_{J},v_{K}]_{_{\VV}}=h\big((v_{I},v_{J})_{_{%
\VV}}\big)\times v_{K},
\end{equation*}%
for any function $h:\mathbb{R}\rightarrow \mathbb{R}$, as the most generic
possibility [ $\mathbb{II}$ ]. However, the tri-linearity of the triple
product $[\cdot ,\cdot ,\cdot ]_{_{\VV}}$ (\ref{triple-VV}) in $\VV$
requires the function $h$ to be linear. Since the constant term of the
linear function $h$ leads to a trivial triple product and is easily refuted
by the other axioms of App. \ref{App-Ind-Axioms}, one can conclude that, up
to an overall (real) factor, (\ref{possib-2}) is the most generic
possibility [ $\mathbb{II}$ ].} (recall definitions (\ref{ind-metric}) and (%
\ref{LL-V-def})):%
\begin{equation}
\LL_{v_{I}v_{J}}\,v_{K}=[v_{I},v_{J},v_{K}]_{_{\VV}}=(v_{I},v_{J})_{_{\VV%
}}\times v_{K}.  \label{possib-2}
\end{equation}%
The triple product $\left[ \cdot ,\cdot ,\cdot \right] _{\VV}$ (\ref%
{triple-VV}) defined by (\ref{possib-2}) satisfies the strong form of axiom (%
$B.\,iii$) and most of other axioms of App. \ref{App-Ind-Axioms}. However,
\textit{at least} within the assumption of \textit{non-degeneracy} of the
metric of the algebra $\NN$ (\textit{cfr.} Footnote 19), it is refuted by
axiom ($B.\,ii$) whenever $\KK$ is larger then a single-generator algebra $%
\KK_{\phi }$. $\blacksquare $
\end{itemize}

This completes the proof of the following\medskip

\textbf{No-Go Theorem}

Assuming the metric of the algebraic system $\NN$ (\ref{NN}) to be \textit{%
non-degenerate} and the \textit{Freudenthal duality }$\mathcal{F}$ to be
defined only for $\NN$ satisfying all the four \textit{FTS} axioms
introduced in Subsec. \ref{Subsec-FTS-def}, then it is \textit{not} possible
to construct a Lagrangian density functional $\mathbf{L}\left[ \Phi (x)%
\right] $ for a $\KK$-valued vector/spinor field $\Phi (x)$ which admits
both (\textit{off-shell}) FTS \textit{gauge} symmetry and (\textit{off-shell}%
) \textit{global} \textit{F-duality }symmetry\textit{\ }$\mathcal{F}$.

\section{\label{FGT-SC}FGT and ($\mathcal{N}=3$, $D=3$) SC CSM Gauge Theories%
}

We will now briefly make some observations on the relation between \textit{%
Freudenthal gauge theory} (FGT) (based on \textit{Freudenthal triple systems}
(\textit{FTS}'s)) and the intense research on triple systems and gauge
theories, in which remarkable advances were achieved after the seminal
papers of Bagger and Lambert \cite{BLG} and Gustavsson \cite{gustavsson}. A
more detailed analysis will be reported in \cite{MST-2}.

Here, we will focus on the relation to superconformal (SC)
Chern-Simons-matter (CSM) gauge theories in $D=3$ (in which the $\mathcal{R}$%
-symmetry structure is richer); we will mainly refer to the mathematical
treatment of \cite{F-1} and \cite{F-2} (see also \cite{F-3}); for an
extensive list of Refs. on BLG theories and related developments, besides
\cite{F-1,F-2,F-3}, we address the reader \textit{e.g.} to the recent
comprehensive review \cite{Lambert-rev}.\smallskip

We anticipate that the symmetry properties (\ref{symm-f-gen}) of the FTS
structure constants on which the FGT is based are generally different from
the ones pertaining to the structure constants on which the BLG-type
theories (such as the ones investigated \textit{e.g.} in \cite%
{Gaiotto-Witten,Palmkvist-Kim}, among others) rely. Among SC CSM $D=3$ gauge
theories, the symmetry (\ref{symm-f-gen}) is indeed consistent \textit{only}
with $\mathcal{N}=3$ (see \textit{e.g.} \cite{F-2}, and Refs. therein).
Disregarding the global (\textit{off-shell}) \textit{Freudenthal duality}, ($%
D=3$) FGT could be viewed as an alternative, purely \textit{bosonic sector}
of the corresponding $\mathcal{N}=3$, $D=3$ SC CSM gauge theory. In fact, as
analyzed in\ Sec. \ref{Subsec-Minimal}, in FGT the non-vanishing of $%
f_{(abcd)}$ allows for terms in the Lagrangian which differ from the usual
ones in BLG theories; for instance, the simplest FGT scalar potential is
\textit{quartic} in the scalar fields (essentially given by $\Delta $ (\ref%
{Delta-def}); see (\ref{FGT-bos-action-min})), whereas in BLG theories it is
of order six (see \textit{e.g.} (19) of \cite{BLG}).\medskip

We start by observing that the set of axioms ($i$), ($iii$) and ($iv$)
defining an FTS\ (as given in Sec. \ref{Subsec-f_abcd}) match the set of
axioms ($a$), ($b$) and ($c$) defining the triple systems based on \textit{%
quaternionic unitary representations} $\mathbf{W}$ of a metric Lie algebra $%
\mathfrak{g}$, as discussed in \cite{F-1} and \cite{F-2} (see \textit{e.g.}
App. A.2.4 of \cite{F-2}, and axioms (125)-(127) therein); in particular,
the FTS axiom ($iii$) is nothing but the so-called \textit{fundamental
identity} of the triple system (see \textit{e.g.} (127) of \cite{F-2}). In
turn, the treatment of \cite{F-1} and \cite{F-2} is based on a construction
due to Faulkner \cite{Faulkner-constr,Faulkner-Ferrar}, which essentially
constructs triple systems from pairs $\left( \mathfrak{g},\mathbf{V}\right) $%
, where $\mathbf{V}$ is a suitable representation\footnote{%
The fourth axiom (\textit{quaternionic condition}; see \textit{e.g.} (128)
of \cite{F-2}) defining Faulkner's triple systems based on $\left( \mathfrak{%
g},\mathbf{W}\right) $ is essentially related to the existence of a
skew-symmetric symplectic invariant bilinear form $\omega $ which raises and
lowers indices.} of $\mathfrak{g}$ \cite{F-1}.

The $\mathfrak{g}$-irreducible decomposition of the rank-$4$ $\mathfrak{g}$%
-invariant structure in $\mathbf{W}$ is given by (124) of \cite{F-2} (also,
\textit{cfr.} Table 2 therein):%
\begin{equation}
S^{2}S^{2}\mathbf{W}\cong S^{4}\mathbf{W}\oplus \mathbf{W}^{(2,2)}.
\label{f}
\end{equation}

In tensor notation, a reformulation\footnote{%
Here, we will not deal with issues of generality of the reformulation (\ref%
{f-2}) of (\ref{f}).} of\ (\ref{f}) reads as follows ($a,b\in \mathbb{R}$):%
\begin{equation}
f_{abcd}=af_{(abcd)}+b\omega _{a(c}\omega _{d)b}.  \label{f-2}
\end{equation}%
(\ref{f-2}) is consistent with the general symmetry of the FTS structure
constants' tensor $f_{abcd}$ given by (\ref{symm-f-gen}); furthermore,
\textit{Freudenthal duality} $\mathcal{F}$ (\ref{dual-field}) can be
consistently introduced whenever $f_{(abcd)}\neq 0$.

It is worth remarking that Brown's definition of Lie algebra $\left(
\mathfrak{g},\mathbf{R}\right) $ \textit{of type} $\mathfrak{e}_{7}$ \cite%
{brown} (\textit{cfr.} ($a$)-($c$) in\ Sec. \ref{Subsec-FD-def}) can be
extended to include also the not completely symmetric part $\omega
_{a(c}\omega _{d)b}$ of (\ref{f-2}) as follows: $\mathbf{R}$ is a
representation space of $\mathfrak{g}$ such that

\begin{description}
\item[($\protect\widehat{a}$)] $\mathbf{R}$ possesses a \textit{%
non-degenerate}, skew-symmetric bilinear $\mathfrak{g}$-invariant form $%
\omega $ (\textit{cfr.} (\ref{sympl-def}) and (\ref{omega-def}));

\item[($\protect\widehat{b}$)] $\mathbf{R}$ possesses a rank-$4$ $\mathfrak{g%
}$-invariant structure $f_{abcd}$ (\ref{f-2}), which allows to define%
\begin{equation}
\widehat{q}\left( x,y,z,w\right) \equiv f_{abcd}x^{a}y^{b}z^{c}w^{d}=2%
\widehat{\Delta }\left( x,y,z,w\right) ;  \label{q-hat-def}
\end{equation}

\item[($\protect\widehat{c}$)] by defining a ternary product $\widehat{%
\mathbf{T}}\left( x,y,z\right) $ on $\mathbf{R}$ as%
\begin{equation}
\left\langle \widehat{\mathbf{T}}\left( x,y,z\right) ,w\right\rangle \equiv
\widehat{q}\left( x,y,z,w\right) ,  \label{deff-hat}
\end{equation}%
then one has%
\begin{equation}
3\left\langle \widehat{\mathbf{T}}\left( x,x,y\right) ,\widehat{\mathbf{T}}%
\left( y,y,y\right) \right\rangle =\left\langle x,y\right\rangle \widehat{q}%
\left( x,y,y,y\right) .  \label{deff-hat-2}
\end{equation}
\end{description}

By enhancing $f_{abcd}=f_{(abcd)}$ to a not completely symmetric $f_{abcd}$
given by (\ref{f-2}), one can conclude that, by virtue of ($\widehat{a}$),
the real parameters $a$ and $b$ can always be chosen such that the inclusion
of $\omega _{a(c}\omega _{d)b}$ in Brown's definition \cite{brown} yields
nothing but an equivalent definition of a Lie algebra \textit{of type} $%
\mathfrak{e}_{7}$; however, as pointed out below, the presence or absence of
the term $\omega _{a(c}\omega _{d)b}$ matters in order to make contact with
\textit{FTS}'s.\medskip

Note that the $\lambda $-dependent $FTS$-defining axiom ($ii$) was not
mentioned so far. However, \textit{at least} for the class of pairs $\left(
\mathfrak{g},\mathbf{R}\right) =\left( \mathfrak{conf}\left( \widehat{%
\mathfrak{J}}\right) ,\mathbf{R}\right) $ reported in Table 1, the
parameters $a$ and $b$ can be fixed consistently with axiom ($ii$), by
further elaborating (\ref{f-2}) as%
\begin{equation}
f_{abcd}=6\lambda f_{(abcd)}-2\lambda \omega _{a(c}\omega _{d)b}.
\label{f-2-lambda}
\end{equation}%
For pairs $\left( \mathfrak{g},\mathbf{R}\right) =\left( \mathfrak{conf}%
\left( \widehat{\mathfrak{J}}\right) ,\mathbf{R}\right) $ with $\mathfrak{g}$
\textit{simple}, both (\ref{f-2-lambda}) and the parameter $\lambda $
acquires a very simple group-theoretical meaning. Indeed, exploiting the
results of \cite{Exc-Reds}, (\ref{f-2-lambda}) can be rewritten as%
\begin{equation}
f_{abcd}=-3\tau f_{(abcd)}+\tau \omega _{a(c}\omega _{d)b}=t_{ab}^{\alpha
}t_{cd}^{\beta }g_{\alpha \beta },  \label{f-2-tau}
\end{equation}%
where $t_{ab}^{\alpha }=t_{\left( ab\right) }^{\alpha }$ is the ($\mathfrak{g%
}$-invariant) realization of the generators of $\mathfrak{g}$ in $\mathbf{R}$%
; the indices $\alpha $ and $a$ respectively are in $\mathbf{Adj}$ and $%
\mathbf{R}$ of $\mathfrak{g}$, whose Cartan-Killing metric is $g_{\alpha
\beta }$. Therefore, $f_{abcd}$ can be defined as the adjoint-trace of the
product of two realizations of generators of $\mathfrak{g}$ in its
representation $\mathbf{R}$. Moreover, the parameter \cite{Exc-Reds}%
\begin{equation}
\tau \equiv \frac{2\text{dim}_{\mathbb{R}}\mathbf{Adj}\left( \mathfrak{g}%
\right) }{\text{dim}_{\mathbb{R}}\mathbf{R}\left( \mathfrak{g}\right) \left(
\text{dim}_{\mathbb{R}}\mathbf{R}\left( \mathfrak{g}\right) +1\right) }%
=-2\lambda  \label{tau-def}
\end{equation}%
expresses the ratio between the sets of indices $\alpha $ and $ab=(ab)$ of $%
t_{ab}^{\alpha }$ (in the treatment above, we set dim$_{\mathbb{R}}\mathbf{R}%
\left( \mathfrak{g}\right) \equiv f$; \textit{cfr.} (\ref{dim-f-def})). By
virtue of the \textit{Gaillard-Zumino embedding} (\ref{GZ-pre}) \cite{GZ}
(or, equivalently of the aforementioned Theorem by Dynkin \cite%
{dynkin-2,Lorente}), $\tau $ expresses the fraction of generators of $%
\mathfrak{sp}\left( f,\mathbb{R}\right) $ which generate its maximal
(generally non-symmetric) sub-algebra $\mathfrak{g}$. Indeed, it holds that%
\begin{equation}
0<\tau \leqslant 1\Leftrightarrow -\frac{1}{2}\leqslant \lambda <0.
\end{equation}

By a suitable generalization of the analysis of \cite{gustavsson-2},
explicitly worked out in \cite{Palmkvist-Kim}, the choice of $f_{abcd}$
given by (\ref{f-2-tau}) can be made also for the pairs $\left( \mathfrak{g},%
\mathbf{R}\right) =\left( \mathfrak{conf}\left( \widehat{\mathfrak{J}}%
\right) ,\mathbf{R}\right) $ with $\mathfrak{g}$ \textit{semi-simple}.
However, in these cases the last step of (\ref{f-2-tau}) does not hold:%
\begin{equation}
f_{abcd}=-3\tau f_{(abcd)}+\tau \omega _{a(c}\omega _{d)b}\neq
t_{ab}^{\alpha }t_{cd}^{\beta }g_{\alpha \beta };
\end{equation}%
in fact, the explicit expression of $t_{ab}^{\alpha }t_{\alpha \mid cd}$ for
these cases has been computed in \cite{Palmkvist-Kim}, and it is such that
\cite{Gaiotto-Witten}%
\begin{equation*}
g_{\alpha \beta }t_{(ab}^{\alpha }t_{c)d}^{\beta }=0.
\end{equation*}%
\medskip

Thus, the \textit{FTS} (the triple system on which the FGT is based) turns
out to be related to the \textit{quaternionic level} of Faulkner's
construction \cite{Faulkner-constr} of triple systems from pairs $\left(
\mathfrak{g},\mathbf{V}\right) $, which has been recently re-analyzed by
\cite{F-1,F-2,F-3} within $D=3$ SC CSM gauge theories.

An important difference with the latter framework is the fact that, in the
treatment of the present paper, \textit{FTS} is defined on the ground field $%
\mathbb{R}$ (recall Footnote 1); this constrains the pair $\left( \mathfrak{g%
},\mathbf{V}\right) =\left( \mathfrak{g},\KK\right) $ such that $\mathbf{V}$
is a \textit{real} representation space of the (\textit{non-compact}) real
algebra $\mathfrak{g}$; some examples, related to conformal symmetries of
\textit{JTS} $\JJ=\widehat{\mathfrak{J}}$, are reported in Table 1. As
mentioned in Sec. \ref{Subsec-Minimal}, we point out that this is not
inconsistent with the physical constraint on matter representations in $D=3$
SC CSM gauge theories; indeed, $\mathbf{V}=\mathbf{W}$ is always assumed to
possess a positive-definite inner product (for \textit{unitarity} of the
corresponding gauge theory), but CS gauge fields are not propagating (and
they are in $\mathbf{Adj}\left( \mathfrak{g}\right) $), and therefore $%
\mathfrak{g}$ does not necessarily have to be endowed with a
positive-definite product, thus allowing for \textit{non-compact} (real)
forms of $\mathfrak{g}$.\medskip

The expression (\ref{f-2}) of the FTS structure constants' tensor $f_{abcd}$
(or, equivalently, for the rank-$4$ $\mathfrak{g}$-invariant structure in $%
\mathbf{W}$ in $\left( \mathfrak{g},\mathbf{V}=\mathbf{W}\right) $-based
Faulkner's construction of triple systems \cite{Faulkner-constr}) entails
two \textquotedblleft extremal" cases:

\begin{enumerate}
\item The case in which $f_{abcd}$ is \textit{completely symmetric} (and
therefore \textit{Freudenthal duality} $\mathcal{F}$ (\ref{dual-field}) can
be consistently introduced). This corresponds to $b=0$ and (up to
redefinition) $a=1$ in (\ref{f-2}):%
\begin{equation}
f_{abcd}=f_{(abcd)},  \label{qLTS}
\end{equation}%
which characterizes Brown's definition \cite{brown} of $\left( \mathfrak{g},%
\mathbf{W}\right) $ as a Lie algebra \textit{of type} $\mathfrak{e}_{7}$ (%
\textit{cfr.} axiom ($b$) in Sec. \ref{Subsec-FD-def}). The corresponding
triple system has been called \textit{quaternionic Lie triple system} (%
\textit{qLTS}) in \cite{F-2}. However, this triple system is \textit{not}
relevant for application to (BLG-type) gauge theories. Indeed, for \textit{%
positive-definite} $\mathbf{W}$ (as assumed for unitarity of the
corresponding gauge theory), $f_{abcd}$ is nothing but the Riemann tensor of
a symmetric hyper-K\"{a}hler manifold, which is \textit{Ricci-flat};
however, any homogeneous Ricci-flat Riemannian manifold is actually \textit{%
Riemann-flat} \cite{Besse,Alekseevsky}. Thus, a \textit{positive-definite} $%
\mathbf{W}$ in \textit{qLTS} (\ref{qLTS}) is necessarily the trivial
representation (\textit{cfr.} Corollary 6 in \cite{F-2}). Remarkably, this
result has a consistent interpretation in the \textit{FTS} framework.
Indeed, it can be checked that (\ref{qLTS}), when plugged into the \textit{%
FTS} axiom ($iii$) (\textit{fundamental identity}) and contracted with $%
x^{a}x^{b}y^{c}y^{e}y^{f}y^{g}$, does \textit{not} yield the axiom ($c$)
which defines a Lie algebra \textit{of type} $\mathfrak{e}_{7}$ \cite{brown}%
. In other words, $\left( \mathfrak{g},\mathbf{W}\right) $ \textit{of type} $%
\mathfrak{e}_{7}$ \cite{brown} is \textit{not} consistent with the \textit{%
FTS} introduced in Secs. \ref{Subsec-FD-def}-\ref{Subsec-f_abcd}; in
particular, the \textit{fundamental identity} ($iii$) is \textit{not}
consistent with axiom ($c$) of Lie algebras \textit{of type} $\mathfrak{e}%
_{7}$ \cite{brown}. As a consequence, the limit of the defining axioms ($i$%
)-($iv$) in which $f_{abcd}$ is taken to be \textit{completely symmetric} (%
\ref{qLTS}) is ill defined; a non-trivial $\lambda \rightarrow 0$ limit in ($%
i$)-($iv$) can still be implemented, but it yields an \textit{FTS} which
does not fulfill the symmetry condition (\ref{qLTS}) \cite{MST-2}.

\item The case in which $f_{abcd}$ lacks its \textit{completely symmetric}
part. This corresponds to $a=0$ and (up to redefinition) $b=1$ in (\ref{f-2}%
):%
\begin{equation}
f_{abcd}=\omega _{a(c}\omega _{d)b}.  \label{aLTS}
\end{equation}%
In this case the \textit{Freudenthal duality} $\mathcal{F}$ (\ref{dual-field}%
) cannot be consistently introduced. The corresponding triple system has
been called \textit{anti-Lie triple system} (\textit{aLTS}) in \cite{F-2};
it characterizes $\mathcal{N}=4$ and $\mathcal{N}=5$ SC CSM gauge theories
in $D=3$, as thoroughly analyzed in \cite{F-2} (see also Table 6 therein),
by elaborating on previous literature (see Refs. therein). A prototypical
case (treated in Example 1 of \cite{Faulkner}) is provided by a consistent
limit of (\ref{f-2-lambda}), given by\footnote{%
Recall that, under the assumption that $\omega $ is non-degenerate, $f$ is
even.} (recall (\ref{dim-f-def})) $\mathfrak{g}=\mathfrak{sp}(f,\mathbb{R})$
and $\mathbf{W}=\mathbf{f}$ (fundamental irrep.). Since%
\begin{equation}
S^{4}\mathbf{f}\equiv \left( \mathbf{f}\times \mathbf{f}\times \mathbf{f}%
\times \mathbf{f}\right) _{s}
\end{equation}%
is irreducible in $\mathfrak{sp}(f,\mathbb{R})$ and contains no singlets, it
follows that $f_{(abcd)}=0$. On the other hand, since $\mathbf{Adj}(%
\mathfrak{sp}(f,\mathbb{R}))=S^{2}\mathbf{f\equiv }\left( \mathbf{f}\times
\mathbf{f}\right) _{s}$, the definition (\ref{tau-def}) also yields $\tau =1$%
, and therefore (\ref{aLTS}) is recovered from (\ref{f-2-lambda}). It is
worth remarking that in this case the resulting \textit{FTS} is not endowed
with a manifestly \textit{JTS}-covariant structure (\ref{FTS}) as in the
original Freudenthal's formulation \cite%
{freudenthal,freudenthal-FTS-1,freudenthal-FTS-2}; the corresponding
(super)gravity theory in $D=4$ can have \textit{at most}\footnote{%
In this theory, the consistency of $\mathcal{N}=1$ local supersymmetry with
a symplectic structure of electric and magnetic fluxes has been studied
\textit{e.g.} in \cite{ADFT-N=1-attractors}; see also \cite%
{Duff-Ferrara-gen-mirror}.} $\mathcal{N}=1$ local supersymmetry, and has a
(non--special) K\"{a}hler scalar coset with algebra $\mathfrak{sp}\left( f,%
\mathbb{R}\right) \ominus \mathfrak{u}(f/2)$ (\textit{upper Siegel half-plane%
}).\medskip
\end{enumerate}

The general triple system under consideration, which interpolates between
\textit{qLTS} (\ref{qLTS}) and \textit{aLTS} (\ref{aLTS}), is endowed with
an $f_{abcd}$ given by (\ref{f-2}) with \textit{both} $a$ and $b$ \textit{%
non-vanishing}. As anticipated, among SC CSM gauge theories in $D=3$, this
is consistent only with $\mathcal{N}=3$ (see \textit{e.g.} \cite{F-2}, and
Refs. therein), which is thus the only amount of (global) supersymmetry for
which \textit{Freudenthal duality} $\mathcal{F}$ (\ref{dual-field}) could
\textit{a priori} be implemented, even if its enforcement as a \textit{global%
} (\textit{off-shell}) symmetry is in contrast with supersymmetry itself, as
implied by the \textit{No-Go theorem} proved in Sec. \ref{Subsec-No--Go}.

It is worth observing that this general case is also consistent with the
\textquotedblleft extension" of the definition of Lie algebras \textit{of
type} $\mathfrak{e}_{7}$ (based on axioms ($\widehat{a}$)-($\widehat{c}$)
above); indeed, up to some redefinitions, the real parameters $a$ and $b$
can always be chosen such that (\ref{f-2}), when plugged into the FTS axiom (%
$iii$) and contracted with $x^{a}x^{b}y^{c}y^{e}y^{f}y^{g}$, does yield the
axiom ($\widehat{c}$) introduced above; the term $\omega _{a(c}\omega _{d)b}$
plays a key role in this result.\bigskip

The above treatment hints for the existence of a class of $\mathcal{N}=3$, $%
D=3$ SC CSM gauge theories in which the gauge Lie algebra and its matter
representation are given by%
\begin{equation}
\left( \mathfrak{g},\mathbf{V}\right) =\left( \mathfrak{conf}\left( \widehat{%
\mathfrak{J}}\right) ,\mathbf{R}\right) ,  \label{algebra-1}
\end{equation}%
namely they are respectively given by the \textit{conformal} symmetries of
rank-$3$, Euclidean Jordan algebras, and by their relevant symplectic
irreps. $\mathbf{R}$, as reported in Table 1.

In this respect, by recalling Sec. \ref{FGT-sugra}, $\mathcal{N}=3$, $D=3$
SC CSM gauge theories based on (\ref{algebra-1}) share the same symmetry
(with different physical meanings) of two other distinct classes of theories
:

\begin{itemize}
\item $D=4$ Maxwell-Einstein (super)gravity theories (ME(S)GT) (with various
amount $\mathcal{N}$ of local supersymmetry) having \textit{symmetric}
scalar manifolds, as discussed in Sec. \ref{FGT-sugra} (and reported in
Table 1);

\item $D=3$ \textit{Freudenthal gauge theories} (FGT's) based on an \textit{%
FTS} $\KK\sim \mathbf{R}\left( \mathfrak{conf}\left( \widehat{\mathfrak{J}}%
\right) \right) $. The consistency of FGT with (global) supersymmetry is an
important difference with respect to $\mathcal{N}=3$ SC CSM gauge theories.
Indeed, the \textit{No-Go Theorem} proved in\ Sec. \ref{Subsec-No--Go}
essentially states that global (\textit{off-shell}) \textit{Freudenthal
duality} is \textit{not} consistent with a non-trivial coupling to
space-time vector/spinor fields, which in turn is a necessary condition for
supersymmetry.
\end{itemize}

These relations among $\mathcal{N}=3$, $D=3$ SC CSM gauge theories, $D=4$
ME(S)GT's and FGT's can actually be extended to the general case in which
the pair $\left( \mathfrak{g},\mathbf{V}=\mathbf{W}\right) $ defines a
generic $FTS$ (based on axioms ($i$)-($iv$)) corresponding, in the sense
outlined above, to the \textit{\textquotedblleft quaternionic level"} of
Faulkner's construction \cite{Faulkner-constr, Faulkner-Ferrar, F-1,F-2,F-3}.

We plan to investigate this interesting interplay of symmetries in future
work \cite{MST-2} (also in view of possible AdS/CFT applications). In
particular, as anticipated above, when disregarding the global (\textit{%
off-shell}) \textit{Freudenthal duality}, it would be interesting to
consider the consistency of ($D=3$) FGT as an alternative, purely \textit{%
bosonic sector} of the corresponding $\mathcal{N}=3$, $D=3$ SC CSM gauge
theory. In fact, as analyzed in\ Sec. \ref{Subsec-Minimal}, in FGT the
non-vanishing of $f_{(abcd)}$ allows for terms in the Lagrangian which
differ from the usual ones in BLG theories; for instance, the simplest FGT
scalar potential is \textit{quartic} in the scalar fields (essentially given
by $\Delta $ (\ref{Delta-def}); see (\ref{FGT-bos-action-min})), whereas in
BLG theories it is of order six (see \textit{e.g.} (19) of \cite{BLG}).

\section{\label{Sec-Conclusion}Concluding Remarks}

In this paper, we have introduced the \textit{Freudenthal Gauge Theory}
(FGT), a gauge theory invariant under two \textit{off-shell} symmetries: a
local, \textit{gauge} symmetry constructed from a \textit{Freudenthal Triple
System} (\textit{FTS}) $\KK$, and a \textit{global} symmetry based on the
so-called \textit{Freudenthal Duality} (\textit{F-duality}) $\mathcal{F}$.

We have presented the most general bosonic action invariant under these two
symmetries, containing a single $\KK$-valued scalar field $\phi (x)$ and a
gauge field $A_{\m}^{ab}(x)\in \KK\otimes _{S}\KK$. The algebraic structure
of the \textit{FTS} ensures that the FGT is well defined and has the
required properties.

One of the building blocks of FGT is the \textit{F-duality }$\mathcal{F}$,
which is a non-linear \textit{anti-involutive} duality ($\mathcal{F}^{2}=-Id$%
) which gives, up to a sign, a one-to-one pairing of elements in $\KK$.

In\ Sec. \ref{Sec-Gen}, we have also analyzed the possibility of
generalizing the simple setup presented in Sec. \ref{Sec-Gauge} by coupling
to space-time vector \textit{and/or} spinor fields, which is a necessary
condition for supersymmetry and is usually a relatively simple step in the
construction of gauge theories. Within the assumption\footnote{%
We leave the possible relaxation of the assumptions on $\mathcal{F}$ \textit{%
and/or} on the metric of the algebraic system to further future
investigation. Concerning the case of \textit{degenerate} metric, see also
Footnote 19.} that \textit{Freudenthal duality }$\mathcal{F}$ can be defined
only for algebraic systems satisfying the \textit{FTS} axioms ($i$)-($iv$)%
\textit{\ }(see Subsec. \ref{Subsec-FTS-def}) we have proved a \textit{No-Go
theorem} (which holds true if the metric of the system is \textit{%
non-degenerate}), which essentially forbids the coupling to space-time
vector \textit{and/or} spinor fields.

However, we point out that such a coupling is possible \textit{at least} if
one relaxes the requirement of invariance under \textit{F-duality}. Despite
the fact that in our treatment there is, \textit{a priori}, no restriction
on the space-time dimension $D$, \textit{non-compact} gauge Lie algebras $%
\mathfrak{g}$ generally yield non-unitary theories in $D\geqslant 4$ (%
\textit{cfr.} the remark below (\ref{FGT-D=4})). However, in $D=3$ this is
no more a problem, and the resulting (\textit{non-Freudenthal-invariant})
FGT can contain both bosonic and fermionic degrees of freedom together with
the Chern-Simons term.

In $D=3$, some intriguing similarities (and important differences) between
FGT and superconformal (SC) Chern-Simons-matter (CSM) gauge theories with $%
\mathcal{N}=3$ global supersymmetry have been discussed in Sec. \ref{FGT-SC}%
. Indeed, among SC CSM gauge theories in $D=3$, a generic \textit{FTS} is
\textit{only} consistent for $\mathcal{N}=3$ (see \textit{e.g.} \cite{F-2},
and Refs. therein), which is thus the only amount of (global) supersymmetry
for which \textit{Freudenthal duality} $\mathcal{F}$ (\ref{dual-field})
could \textit{a priori} be implemented, even if its enforcement as a \textit{%
global} (\textit{off-shell}) symmetry is in contrast with supersymmetry
itself, as implied by the \textit{No-Go theorem} proved in Sec. \ref%
{Subsec-No--Go}.

It is worth recalling here that our treatment hints for the existence of a
class of $\mathcal{N}=3$, $D=3$ SC CSM gauge theories in which the gauge Lie
algebra is given by (\ref{algebra-1}), namely by the \textit{conformal}
algebras $\mathfrak{g}=\mathfrak{conf}\left( \widehat{\mathfrak{J}}\right) $
of rank-$3$, Euclidean Jordan algebras, and by their relevant symplectic
irreps. $\mathbf{R}$, as reported in Table 1. In this respect, such $%
\mathcal{N}=3$, $D=3$ SC CSM gauge theories share the same symmetry (with
different physical meanings) of two other distinct classes of theories :
\textbf{I}] $D=4$ Maxwell-Einstein (super)gravity theories (ME(S)GT) (with
various amount $\mathcal{N}$ of local supersymmetry) with \textit{symmetric}
scalar manifolds, as discussed in Sec. \ref{FGT-sugra} (and reported in
Table 1); \textbf{II}] $D=3$ FGT's based on an \textit{FTS} $\KK\sim \mathbf{%
R}\left( \mathfrak{conf}\left( \widehat{\mathfrak{J}}\right) \right) $.

These relations among $\mathcal{N}=3$, $D=3$ SC CSM gauge theories, $D=4$
ME(S)GT's and $D=3$ FGT's can actually be extended to the general case in
which the pair $\left( \mathfrak{g},\mathbf{V}=\mathbf{W}\right) $ defines a
generic $FTS$ (based on axioms ($i$)-($iv$)) corresponding, as discussed in
Sec. \ref{FGT-SC}, to the \textit{\textquotedblleft quaternionic level"} of
Faulkner's construction \cite{Faulkner-constr, Faulkner-Ferrar, F-1,F-2,F-3}.

We plan to investigate this interesting interplay of symmetries in future
work \cite{MST-2} (also in view of possible AdS/CFT applications). In
particular, when disregarding the global (\textit{off-shell}) \textit{%
Freudenthal duality}, it will be interesting to consider the consistency of $%
D=3$ FGT as an alternative, purely \textit{bosonic sector} of the
corresponding $\mathcal{N}=3$, $D=3$ SC CSM gauge theory. In fact, as
analyzed in\ Sec. \ref{Subsec-Minimal}, in FGT the non-vanishing of $%
f_{(abcd)}$ allows for terms in the Lagrangian which differ from the usual
ones in BLG theories; for instance, the simplest FGT scalar potential is
\textit{quartic} in the scalar fields (essentially given by $\Delta $ (\ref%
{Delta-def}); see (\ref{FGT-bos-action-min})), whereas in BLG theories it is
of order six (see \textit{e.g.} (19) of \cite{BLG}).

The close relation between the particular class $\KK\left( \widehat{%
\mathfrak{J}}\right) $ of \textit{FTS}'s and exceptional Lie algebras $%
\mathfrak{g}$ (discussed in Secs. \ref{Subsec-Lie-Algebras} and \ref%
{Gauge-Algebras-type-e7}) could also be used to investigate the possible
relation (\textit{duality?}) between FGT and Yang-Mills gauge theory with
exceptional gauge Lie algebra $\mathfrak{g}$. This is certainly possible,
but one should recall that exceptional Lie groups cannot be embedded into
standard matrix groups, and thus the resulting Yang-Mills theory would not
have the standard Maxwell term constructed from trace over matrices.
Geometrically, a better way to understand this model is by noting that the
exceptional Lie groups can be embedded as matrix groups over octonions $%
\mathbb{O}$ \cite{Baez}; thus, the $\KK\left( \widehat{\mathfrak{J}}\right) $%
-based FGT would be \textit{dual} to a standard Yang-Mills theory over%
\footnote{%
For similar formulations, see \textit{e.g.} \cite{Yamazaki,Castro1,Castro2},
and Refs. therein.} $\mathbb{O}$.

The present investigation proved the \textit{quartic} polynomial $\Delta $ (%
\ref{Delta-def}) to be \textit{invariant} not only under \textit{Freudenthal
duality} $\mathcal{F}$ (\ref{dual-field}), but also under the (\textit{global%
} or \textit{gauged}) transformation based on the \textit{FTS} triple
product (\ref{T-def}). It will be interesting to investigate the physical
meaning of such an invariance of $\Delta $ \textit{e.g.} within black hole
physics \cite{duff} and flux compactifications \cite{Cassani:2009na}, in
which $\Delta $ occurs in relation respectively to the Bekenstein-Hawking
\cite{BH-1,BH-2} black hole entropy and to the cosmological constant.
Interesting recent advances on Freudenthal duality \cite{Levay,FDL} might
also lead to further developments in understanding FGT.

Finally, we would like to point out that the \textit{FTS} has another
intriguing geometrical interpretation in terms of the so-called \textit{%
metasymplectic geometry,} introduced decades ago by Freudenthal \cite%
{freudenthal} \cite{LandsbergManivel}. In such a geometric framework, two
points can define, instead of a line passing through them as in the standard
geometry, two more relations, called \textit{interwoven} and \textit{hinged}%
. Furthermore, to each set of points there corresponds a set of \textit{dual}
geometrical objects called \textit{symplecta}, satisfying relations which
are \textit{dual} to the aforementioned three ones among the points. In this
bizarre geometrical setup, the \textit{FTS} axioms acquire a natural
geometrical interpretation, and the relation to the exceptional Lie algebras
becomes more transparent. We leave the possible physical interpretation of
such a fascinating geometry within FGT for further future investigation.


\section*{Acknowledgements}

We are grateful to Raymond Stora for encouragement, enlightening discussions
and careful reading of the manuscript.

A.M. would like to thank Rob Knoops for discussions.

A.M. would also like to thank the Department of Physics, University of
California at Berkeley, where part of this project was done, for kind
hospitality and stimulating environment.

The work of B. Z. ~has been supported in part by the Director, Office of
Science, Office of High Energy and Nuclear Physics, Division of High Energy
Physics of the U.S. Department of Energy under Contract No.
DE-AC02-05CH11231, and in part by NSF grant 30964-13067-44PHHXM.


\appendix
\appendixpage

\section{\label{App-FD}Freudenthal Duality}

In this Appendix, generalizing the treatment of \cite{duff} (in turn
referring to \cite{brown}) to a generic \textit{FTS} $\KK$ (see also \cite%
{alessio}), we present the proof that the \textit{quartic} polynomial $%
\Delta (\phi )$ (\ref{Delta-def}) is \textit{invariant} under the \textit{%
Freudenthal duality }$\mathcal{F}$ (\ref{dual-field}).

By recalling definition (\ref{LL-def}), we can restate the derivation
property (\textit{FTS} axiom ($iii^{\prime }$)) as follows:
\begin{equation}
[\LL_{\phi_L\phi_M} , \LL_{\phi_I\phi_J}] \phi_K = \LL_{(\phi_L\phi_M\phi_I)%
\phi_J}\phi_K + \LL_{\phi_I(\phi_L\phi_M\phi_J)}\phi_K.
\end{equation}
Since this equation is true for any element $\phi_K\in\KK$, it is true as an
operator equation for $\LL_{\phi_I\phi_J}$. Setting $I=J=L=M$, we find that
\begin{equation}
[\LL_{\phi\phi},\LL_{\phi\phi}] = \LL_{T(\phi)\phi} + \LL_{\phi T(\phi)} = 2%
\LL_{\phi T(\phi)}
\end{equation}

%
%
%
%
%
%
%
%
%
%
\noindent where the FTS axiom ($i$) of Subsec. \ref{Subsec-FTS-def} has been
used. Since the commutator of an operator with itself must vanish, the above
equation must be equal to zero:
\begin{equation}
\LL_{\phi T(\phi)} = 0
\end{equation}
This means, again by the derivation property of $\LL$, that both $\LL%
_{T(\phi )\phi }$ and $\LL_{\phi T(\phi )}$ act like \textit{annihilation
operators} on any element $\phi_K\in \KK$.

Then, by recalling the definition (\ref{Delta-def}), from the \textit{FTS}
axiom ($ii$) of Subsec. \ref{Subsec-FTS-def} one obtains:
\begin{align}
\LL_{T(\phi )T(\phi )}\phi & =T(\phi )T(\phi )\phi  \notag \\
& =T(\phi )\phi \,T(\phi )+2\lam\,\la T(\phi ),\phi \ra\,T(\phi )+\lam\,\la %
T(\phi ),\phi \ra\,T(\phi )-\lam\,\la T(\phi ),T(\phi )\ra\,\phi  \notag \\
& =6\lam\,\Delta (\phi )\,T(\phi );  \label{res-1} \\
\LL_{\phi \phi }T(\phi )& =\phi \phi \,T(\phi )=-6\lam\,\Delta (\phi )\phi .
\label{res-2}
\end{align}%
%
%
%
%
%
%
%
%
%
%
%
%
%
Consequently, the direct evaluation of $T(T(\phi ))$ reads:
\begin{align}
T(T(\phi ))& =\LL_{T(\phi )T(\phi )}T(\phi )=6\lam\,\Delta (\phi )\Big(%
T(\phi )\phi \phi +\phi \,T(\phi )\phi +\phi \phi \,T(\phi )\Big)  \notag \\
& =-\big(6\lam\,\Delta (\phi )\big)^{2}\phi .  \label{res-3}
\end{align}%
From result (\ref{res-3}), by assuming $6\lam\,\Delta (\phi )\neq 0$ (see
discussion in Subsec. \ref{Subsec-FD-def}, in particular point (\textbf{III}%
)), one can check that the following two statements hold true:

\begin{enumerate}
\item The \textit{Freudenthal duality }$\mathcal{F}$ (\ref{dual-field}) is
an \textit{anti-involution} in the \textit{FTS} $\KK$, namely it squares to
negative identity (\textit{cfr.} (\ref{anti-involutivity}) and point (%
\textbf{I}) of Subsec. \ref{Subsec-FD-def}):%
\begin{equation}
\mathcal{F}^{2}\equiv \mathcal{F}\circ \mathcal{F}=-Id.
\label{anti-involutivity-2}
\end{equation}

\item The \textit{quartic} polynomial $\Delta (\phi )$ (\ref{Delta-def}) is
\textit{invariant} under the \textit{Freudenthal duality }$\mathcal{F}$ (\ref%
{dual-field}), namely (\textit{cfr.} (\ref{Delta-inv}))%
\begin{equation*}
\Delta (\phi )=\Delta (\tilde{\phi}),~~q.e.d.~\blacksquare
\end{equation*}
\end{enumerate}

\section{\label{App-Symm-Scalar-Kinetic}Space-Time Symmetry of Scalar
Kinetic Term}

In order to prove the symmetry (\ref{symm-scalar-kin}) of the FGT kinetic
scalar term under the exchange of its space-time indices, one needs to
re-write it only in terms of the $\KK$-valued scalar field $\phi (x)$, by
recalling the definitions (\ref{Delta-def}) and (\ref{dual-field}) of the
\textit{quartic} polynomial $\Delta \left( \phi \right) $ and of \textit{%
F-dual} field $\widetilde{\phi }(x)$.

One starts by computing the \textit{FTS} gauge covariant derivative of $%
\widetilde{\phi }(x)$, as follows:
\begin{align}
D_{\mu }\tilde{\phi}(x)& =\text{sgn}\left( \Delta (\phi )\right) \frac{1}{%
\sqrt{6}}D_{\mu }\left( \frac{T(\phi )}{\sqrt{\,\left\vert \lam\Delta (\phi
)\right\vert }}\right)  \notag \\
& =\frac{\text{sgn}\left( \Delta (\phi )\right) }{\sqrt{6|\lam\Delta (\phi )|%
}}\left[ 3\LL_{\phi \phi }D_{\mu }\phi +6\lam\la D_{\mu }\phi ,\phi \ra\phi +%
\frac{\la D_{\mu }\phi ,T(\phi )\ra}{\Delta (\phi )}T(\phi )\right]
\label{ress-1}
\end{align}%
As an aside, notice that the $\Delta (\phi )$ in the denominator of the last
term does not have absolute signs attached to it. Plugging this expression
into the kinetic term (prior to contraction with $\eta ^{\mu \nu }$) yields
its following explicit re-writing only in terms of $\phi (x)$:
\begin{align}
\frac{1}{2}\la D_{\mu }\phi ,D_{\nu }\tilde{\phi}\ra& =\frac{\text{sgn}%
\left( \Delta (\phi )\right) }{2\sqrt{6|\lam\Delta (\phi )|}}\bigg[3\la %
D_{\mu }\phi ,\LL_{\phi \phi }D_{\nu }\phi \ra+6\lam\la D_{\mu }\phi ,\phi %
\ra\la D_{\nu }\phi ,\phi \ra  \notag \\
& +\frac{1}{\Delta (\phi )}\la D_{\mu }\phi ,T(\phi )\ra\la D_{\nu }\phi
,T(\phi )\ra\bigg].  \label{ress-2}
\end{align}%
On the other hand, the second and third term of (\ref{ress-2}) are
manifestly symmetric under $\mu \leftrightarrow \nu $, the symmetry of the
first term can be proved as follows:
\begin{equation}
\la D_{\mu }\phi ,\LL_{\phi \phi }D_{\nu }\phi \ra=-\la\LL_{\phi \phi
}D_{\mu }\phi ,D_{\nu }\phi \ra=\la D_{\nu }\phi ,\LL_{\phi \phi }D_{\mu
}\phi \ra,
\end{equation}%
thus implying the result (\ref{symm-scalar-kin}). $\blacksquare $

\section{\label{App-Ind-Axioms}Axioms of $\VV$}

As discussed in Subsec. \ref{Subsec-No--Go}, we report here the five axioms
induced on $\VV$ by the five axioms ($o$)-($iv$) of the algebra $\NN$ (in
addition to the ones already introduced on $\VV$ for other physical reasons,
such as the ones required by the Bose \textit{and/or} Fermi statistics for
the fields $v_{I}\in $ $\VV$). In particular, in the proof of the \textbf{%
No-Go Theorem} in Subsec. \ref{Subsec-No--Go}, a crucial role is played by
axioms ($B.\,iii$) and ($B.\,ii$).

\begin{itemize}
\item[($B.\,o$)] $(v_{I},v_{J})_{_{\VV}}=(v_{J},v_{I})_{_{\VV}};$

\item[($B.\,i$)] $[v_{I},v_{J},v_{K}]_{_{\VV}}=[v_{J},v_{I},v_{K}]_{_{\VV}};$

\item[($B.\,ii$)] $(\phi _{I}\phi _{J}\phi _{K})\otimes \Big(\lbrack
v_{I},v_{J},v_{K}]_{_{\VV}}-[v_{I},v_{K},v_{J}]_{_{\VV}}\Big)$\newline
\hspace*{1cm}$=\la\phi _{J},\phi _{K}\ra\,\phi _{I}\otimes \Big(2\m%
\,(v_{J},v_{K})_{_{\VV}}\times v_{I}-2\lam\,[v_{I},v_{J},v_{K}]_{_{\VV}}%
\Big)
$\newline
\hspace*{1.2cm}$+\la\phi _{I},\phi _{K}\ra\,\phi _{J}\otimes \Big(\m%
\,(v_{I},v_{K})_{_{\VV}}\times v_{J}-\lam\,[v_{I},v_{J},v_{K}]_{_{\VV}}\Big)$%
\newline
\hspace*{1.2cm}$-\la\phi _{I},\phi _{J}\ra\,\phi _{K}\otimes \Big(\m%
(v_{I},v_{J})_{_{\VV}}\times v_{K}-\lam\,[v_{I},v_{J},v_{K}]_{_{\VV}}\Big);$

\item[($B.\,iii$)] $0=(\phi _{L}\phi _{M}\phi _{I})\phi _{J}\phi _{K}\otimes %
\Big(\big[v_{L},v_{M},[v_{I},v_{J},v_{K}]_{_{\VV}}\big]_{_{\VV}}-\big[%
\lbrack v_{L},v_{M},v_{I}]_{_{\VV}},v_{J},v_{K}\big]_{_{\VV}}\Big)$\newline
\hspace*{0.6cm}$+\phi _{I}(\phi _{L}\phi _{M}\phi _{J})\phi _{K}\otimes \Big(%
\big[v_{L},v_{M},[v_{I},v_{J},v_{K}]_{_{\VV}}\big]_{_{\VV}}-\big[%
v_{I},[v_{L},v_{M},v_{J}]_{_{\VV}},v_{K}\big]_{_{\VV}}\Big)$\newline
\hspace*{0.6cm}$+\phi _{I}\phi _{J}(\phi _{L}\phi _{M}\phi _{K})\otimes \Big(%
\big[v_{L},v_{M},[v_{I},v_{J},v_{K}]_{_{\VV}}\big]_{_{\VV}}-\big[%
v_{I},v_{J},[v_{L},v_{M},v_{K}]_{_{\VV}}\big]_{_{\VV}}\Big);$

\item[($B.\,iv$)] $\Big(\lbrack v_{L},v_{M},v_{I}]_{_{\VV}},v_{J}\Big)_{_{\VV%
}}+\Big(v_{I},[v_{L},v_{M},v_{J}]_{_{\VV}}\Big)_{_{\VV}}=0.$
\end{itemize}


\begin{thebibliography}{99}
\bibitem{nambu} Y.~Nambu, \textit{Generalized Hamiltonian dynamics,} Phys.\
Rev.\ D \textbf{7}, 2405 (1973).

\bibitem{takhtajan} R. Chatterjee and L. Takhtajan, \textit{Aspects of
Classical and Quantum Nambu Mechanics}, Lett. Math. Phys. \textbf{37}, 475
(1996), \texttt{hep-th/9507125}.

\bibitem{G1} M. G\"{u}naydin, \textit{Quadratic Jordan formulation of
quantum mechanics and construction of Lie (super)algebras from Jordan
(super)algebras}, Ann. Israel Phys. Soc. no. 3, (1980) 279. Presented at 8th
Int. Colloq. on Group Theoretical Methods in Physics, Kiriat Anavim, Israel,
March 25-29, 1979.

\bibitem{G2} M. G\"{u}naydin, \textit{The exceptional superspace and the
quadratic Jordan formulation of quantum mechanics}, in : \textit{%
\textquotedblleft Elementary particles and the universe: Essays in honor of
Murray Gell-Mann"}, Pasadena 1989, pp. 99-119 (J. Schwarz Editor), Cambridge
University Press.

\bibitem{G3} M. G\"{u}naydin, C. Piron, and H. Ruegg, \textit{Moufang Plane
and Octonionic Quantum Mechanics}, Commun. Math. Phys. \textbf{61}, 69
(1978).

\bibitem{G4} I. Bars and M. G\"{u}naydin, \textit{Construction of Lie
Algebras and Lie Superalgebras from Ternary Algebras}, J. Math.Phys. \textbf{%
20}, 1977 (1979).

\bibitem{G5} I. Bars and M. G\"{u}naydin, \textit{Dynamical Theory of
Subconstituents based on Ternary Algebras}, Phys. Rev. \textbf{D22}, 1403
(1980) 1403.

\bibitem{G6} M. G\"{u}naydin and S. Hyun, \textit{Ternary algebraic approach
to extended superconformal algebras}, Nucl. Phys. \textbf{B373}, 688 (1992).

\bibitem{G7} M. G\"{u}naydin, \textit{Extended superconformal symmetry,
Freudenthal triple systems and gauged WZW models}, \texttt{%
arXiv:hep-th/9502064 [hep-th]}. Presented at the Gursey Memorial Conference
I: On Strings and Symmetries, June 6-10 1994, Istanbul, Turkey.

\bibitem{BLG} J. Bagger and N. Lambert, \textit{Gauge symmetry and
supersymmetry of multiple }$\mathit{M2}$\textit{-branes}, Phys. Rev. \textbf{%
D77} (2008), 065008, \texttt{arXiv:0711.0955 [hep-th]}.

\bibitem{gustavsson} A. Gustavsson, \textit{Algebraic structures on parallel
M2-branes}, Nucl.Phys. \textbf{B811} (2009), 66--76, \texttt{arXiv:0709.1260
[hep-th]}.

\bibitem{Lambert-rev} J. Bagger, N. Lambert, S. Mukhi, and C. Papageorgakis,
\textit{Membranes in }$\mathit{M}$\textit{-theory}, \texttt{arXiv:1203.3546
[hep-th]}.

\bibitem{ho} P. M. Ho, Y. Imamura, Y. Matsuo, and S. Shiba, $\mathit{M5}$%
\textit{-brane in three-form flux and multiple }$\mathit{M2}$\textit{-branes}%
, JHEP \textbf{0808}, 014 (2008), \texttt{arXiv:0805.2898 [hep-th]}.

\bibitem{Gunaydin:1983rk} M. G\"{u}naydin, G. Sierra, and P. K. Townsend,
\textit{Exceptional Supergravity Theories and the Magic Square}, Phys.Lett.
\textbf{B133} (1983), 72.

\bibitem{Gunaydin:1983bi} M. G\"{u}naydin, G. Sierra, and P. K. Townsend,
\textit{The Geometry of }$\mathcal{N}\mathit{=2}$\textit{\ Maxwell-Einstein
Supergravity and Jordan Algebras}, Nucl.Phys. \textbf{B242} (1984), 244.

\bibitem{G8} M. G\"{u}naydin, K. Koepsell, and H. Nicolai, \textit{Conformal
and quasiconformal realizations of exceptional Lie groups}, Commun. Math.
Phys. \textbf{221}, 57 (2001), \texttt{hep-th/0008063}.

\bibitem{G9} M. G\"{u}naydin and O. Pavlyk, \textit{Spectrum Generating
Conformal and Quasiconformal }$\mathit{U}$\textit{-Duality Groups,
Supergravity and Spherical Vectors}, JHEP \textbf{1004} (2010) 070, \texttt{%
arXiv:0901.1646 [hep-th]}.

\bibitem{GP} M. G\"{u}naydin and O. Pavlyk, \textit{Quasiconformal
Realizations of E}$_{6(6)}$\textit{, E}$_{7(7)}$\textit{, E}$_{8(8)}$\textit{%
\ and SO(n+3,m+3)}, $\mathcal{N}\geqslant $\textit{4 Supergravity and
Spherical Vectors}, Adv. Theor. Math. Phys. \textbf{13} (2009), \texttt{%
arXiv:0904.0784 [hep-th]}.

\bibitem{G-Lects} M. G\"{u}naydin, \textit{Lectures on Spectrum Generating
Symmetries and U-duality in Supergravity, Extremal Black Holes, Quantum
Attractors and Harmonic Superspace}, Springer Proc. Phys. \textbf{134}
(2010), \texttt{arXiv:0908.0374 [hep-th]}.

\bibitem{Borsten:2011ai} L. Borsten, M. J. Duff, S. Ferrara, A. Marrani, W.
Rubens, \textit{Explicit Orbit Classification of Reducible Jordan Algebras
and Freudenthal Triple Systems}, \texttt{arXiv:1108.0908 [math.RA]}.

\bibitem{Borsten:2011nq} L. Borsten, M. J. Duff, S. Ferrara, A. Marrani, W.
Rubens, \textit{Small Orbits}, Phys. Rev. \textbf{D85} (2012) 086002,
\texttt{arXiv:1108.0424 [hep-th]}.

\bibitem{brown} R. B. Brown, \textit{Groups of type E}$_{7}$, J. Reine
Angew. Math. \textbf{236} (1969), 79--102.

\bibitem{duff} L. Borsten, D. Dahanayake, M.J. Duff, and W. Rubens, \textit{%
Black holes admitting a Freudenthal dual}, Phys.Rev. D80 (2009), 026003,
\texttt{arXiv:0903.5517 [hep-th]}.

\bibitem{alessio} S. Ferrara, A. Marrani, and A. Yeranyan, \textit{%
Freudenthal Duality and Generalized Special Geometry}, Phys. Lett. \textbf{%
B701} (2011), 640--645, \texttt{arXiv:1102.4857 [hep-th]}.

\bibitem{Ceresole:2011xd} A. Ceresole, S. Ferrara, A. Marrani, and A.
Yeranyan, \textit{Small Black Hole Constituents and Horizontal Symmetry},
JHEP \textbf{1006}, 078 (2011), \texttt{arXiv:1104.4652 [hep-th]}.

\bibitem{Ferrara:2011aa} S. Ferrara and A. Marrani, \textit{Black Holes and
Groups of Type }$\mathit{E}_{7}$, \texttt{arXiv:1112.2664 [hep-th]}.

\bibitem{Ferrara:2011dz} S. Ferrara and R. Kallosh, \textit{Creation of
Matter in the Universe and Groups of Type }$\mathit{E}_{7}$, JHEP \textbf{%
1112}, 096 (2011), \texttt{arXiv:1110.4048 [hep-th]}.

\bibitem{Ferrara:2012qp} S. Ferrara, R. Kallosh, and A. Marrani, \textit{%
Degeneration of Groups of Type }$\mathit{E}_{7}$\textit{\ and Minimal
Coupling in Supergravity}, JHEP \textbf{1206}, 074 (2012), \texttt{%
arXiv:1202.1290 [hep-th]}.

\bibitem{CJ-1} E. Cremmer and B. Julia, \textit{The }$\mathcal{N}\mathit{=8}
$\textit{\ Supergravity Theory. 1. The Lagrangian}, Phys. Lett. \textbf{B80}%
, 48 (1978). E. Cremmer and B. Julia, \textit{The }$\mathit{SO(8)}$\textit{\
Supergravity}, Nucl. Phys. \textbf{B159}, 141 (1979).

\bibitem{HT-1} C. Hull and P. K. Townsend, \textit{Unity of Superstring
Dualities}, Nucl. Phys. \textbf{B438}, 109 (1995), \texttt{hep-th/9410167}.

\bibitem{BH-1} J. D. Bekenstein, \textit{Black holes and entropy}, Phys.
Rev. \textbf{D7} (1973), 2333--2346.

\bibitem{BH-2} S.W. Hawking, \textit{Gravitational radiation from colliding
black holes}, Phys. Rev. Lett. \textbf{26} (1971), 1344--1346.

\bibitem{MST-2} \textit{More on Freudenthal Gauge Theory and Jordan Algebras}%
, to appear (2012).

\bibitem{Faulkner-constr} J. R. Faulkner, \textit{On the Geometry of Inner
Ideals}, J. Algebra \textbf{26}, 1 (1973).

\bibitem{F-1} P. de Medeiros, J. Figueroa-O'Farrill, E. Mendez-Escobar, and
P. Ritter, \textit{On the Lie-Algebraic Origin of Metric 3-Algebras},
Commun. Math. Phys. \textbf{290}, 871 (2009), \texttt{arXiv:0809.1086
[hep-th]}.

\bibitem{F-2} P. de Medeiros, J. Figueroa-O'Farrill, and E. Mendez-Escobar,
\textit{Superpotentials for Superconformal Chern-Simons Theories from
Representation Theory}, J. Phys. \textbf{A42}, 485204 (2009), \texttt{%
arXiv:0908.2125 [hep-th]}.

\bibitem{freudenthal} H. Freudenthal, \textit{Beziehungen der E7 und E8 zur
Oktavenebene. II}, Nederl. Akad. Wetensch. Proc. Ser. A. \textbf{57} (1954),
363--368 = Indag. Math. \textbf{16}, 363--368 (1954).

\bibitem{freudenthal-FTS-1} H. Freudenthal, \textit{Oktaven, ausnahmegruppen
und oktavengeometrie}, Geom. Dedicata 19 (1985), 7.

\bibitem{freudenthal-FTS-2} K. McCrimmon, \textit{The
freudenthal-springer-tits construction of exceptional jordan algebras},
Trans. Amer. Math. Soc. \textbf{139} (1969), 495--510.

\bibitem{Faulkner} J. R. Faulkner, \textit{A construction of Lie algebras
from a class of ternary algebras}, Trans. Amer. Math. Soc. \textbf{155}
(1971), 397--408.

\bibitem{McCrimmon} K. McCrimmon, \textit{A Taste of Jordan Algebras},
Springer-Verlag New York Inc., New York, 2004.

\bibitem{yamaguti} K. Yamaguti and H. Asano, \textit{On the Freudenthal's
construction of exceptional Lie algebras}, Proc. Japan Acad. \textbf{51}
(1975), no. 4, 253--258.

\bibitem{GZ} M. K. Gaillard and B. Zumino, \textit{Duality Rotations for
Interacting Fields}, Nucl. Phys. \textbf{B193} (1981), 221.

\bibitem{dynkin-2} E. B. Dynkin, \textit{The Maximal Subgroups of the
Classical Groups}, American Mathematical Society Translations Series 2, vol.
\textbf{6} (1957), 245 -- 378.

\bibitem{Lorente} M. Lorente and B. Gruber, \textit{Classification of
Semisimple Subalgebras of Simple Lie Algebras}, J. Math. Phys. \textbf{13},
1639 (1972).

\bibitem{ferrar} J. C. Ferrar, \textit{Strictly regular elements in
Freudenthal triple systems}, Trans. Amer. Math. Soc. \textbf{174} (1972),
313--331 (1973).

\bibitem{Okubo-1} S. Okubo, \textit{Triple Products and Yang-Baxter
Equation. 1. Octonionic and Quaternionic Triple Systems}, J. Math. Phys.
\textbf{34}, 3273 (1993), \texttt{hep-th/9212051}.

\bibitem{Okubo-2} S. Okubo, \textit{Triple Products and Yang-Baxter
Equation. 2. Orthogonal and Symplectic Triple Systems}, J. Math. Phys.
\textbf{34}, 3292 (1993), \texttt{hep-th/9212052}.

\bibitem{G11} I. L. Kantor, \textit{Certain generalizations of Jordan
algebras}, Trudy Sem. Vektor. Tenzor. Anal. \textbf{16}, 407 (1972).

\bibitem{Palmkvist} J. Palmkvist, \textit{A Realization of the Lie algebra
associated to a Kantor triple system}, J. Math. Phys. \textbf{47}, 023505
(2006), \texttt{math/0504544}.

\bibitem{G12} I. Kantor and I. Skopets, \textit{Some results on Freudenthal
triple systems}, Sel. Math. Sov. \textbf{2}, 293 (1982).

\bibitem{ADFT-N=1-attractors} L. Andrianopoli, R. D'Auria, S. Ferrara and M.
Trigiante, \textit{Black-hole attractors in }$\mathcal{N}=1$\textit{\
supergravity}, JHEP 0707 (2007) 019, \texttt{hep-th/0703178 [hep-th]}.

\bibitem{SAM-Lectures} S. Bellucci, S. Ferrara, M. G\"{u}naydin, and A.
Marrani, \textit{SAM Lectures on Extremal Black Holes in d=4 Extended
Supergravity}, Springer Proc. Phys. \textbf{134} (2010), 1--30, \texttt{%
arXiv:0905.3739 [hep-th]}.

\bibitem{Manivel-1} L. Manivel, \textit{Configurations of lines and models
of Lie algebras}, Journal of Algebra 304, Vol. 1 (2006), 457 -- 486, \texttt{%
arXiv:math/0507118}.

\bibitem{N=5} B. de Wit and H. Nicolai, \textit{Extended Supergravity with
Local }$\mathit{SO(5)}$\textit{\ Invariance}, Nucl. Phys. \textbf{B188}
(1981), 98.

\bibitem{Tollsten} B. De Wit, H. Nicolai, and H. K. Tollsten, \textit{%
Locally supersymmetric }$\mathit{D=3}$\textit{\ nonlinear sigma models},
Nucl. Phys. \textbf{B392} (1993) 3-38, \texttt{hep-th/9208074}.

\bibitem{Duff:1995sm} M.J. Duff, J. T. Liu, and J. Rahmfeld, \textit{%
Four-dimensional string-string-string triality}, Nucl. Phys. \textbf{B459}
(1996), 125-129, \texttt{hep-th/9508094}.

\bibitem{Behrndt:1996hu} K. Behrndt, R. Kallosh, J. Rahmfeld, M. Shmakova,
and W. K. Wong, \textit{STU black holes and string triality}, Phys. Rev.
\textbf{D54} (1996), 6293--6301, \texttt{hep-th/9608059}.

\bibitem{Andrianopoli:1997pn} L. Andrianopoli, R. D'Auria, and S. Ferrara, $%
\mathit{U}$\textit{\ invariants, black hole entropy and fixed scalars},
Phys.Lett. \textbf{B403} (1997), 12--19, \texttt{hep-th/9703156}.

\bibitem{Ferrara:2008ap} S. Ferrara, A. Gnecchi, and A. Marrani, $\mathit{d=4%
}$\textit{\ Attractors, Effective Horizon Radius and Fake Supergravity},
Phys.Rev. \textbf{D78} (2008), 065003, \texttt{arXiv:0806.3196 [hep-th]}.

\bibitem{Roest:2009sn} D. Roest and H. Samtleben, \textit{Twin Supergravities%
}, Class. Quant.Grav. \textbf{26} (2009), 155001, \texttt{arXiv:0904.1344
[hep-th]}.

\bibitem{G10} M. G\"{u}naydin, S. McReynolds and M. Zagermann, \textit{The }$%
\mathit{R}$\textit{-map and the coupling of }$\mathcal{N}\mathit{=2}$\textit{%
\ tensor multiplets in 5 and 4 dimensions}, JHEP \textbf{0601} (2006) 168,
\texttt{arXiv:hep-th/0511025}.

\bibitem{NML} M. Henneaux and C. Teitelboim, \textit{Dynamics of chiral
(selfdual) }$\mathit{p}$\textit{-forms}, Phys. Lett. \textbf{B206}, 650
(1988). J. H. Schwarz and A. Sen, \textit{Duality symmetric actions}, Nucl.
Phys. \textbf{B411}, 35 (1994), \texttt{hep-th/9304154}. C. Hillmann, $%
E_{7(7)}$\textit{\ invariant Lagrangian of }$d=4$\textit{\ }$\mathcal{N}=8$%
\textit{\ supergravity}, JHEP \textbf{1004}, 010 (2010), \texttt{0911.5225
[hep-th]}.

\bibitem{DF} E. Cremmer, B. Julia, H. Lu, and C. Pope, \textit{Dualization
of dualities. 1.}, Nucl. Phys. \textbf{B523} (1998) 73, \texttt{%
arXiv:hep-th/9710119 [hep-th]}.

\bibitem{FDL} L. Borsten, M. J. Duff, S. Ferrara and A. Marrani, \textit{%
Freudenthal Dual Lagrangians}, \texttt{arXiv:1212.3254 [hep-th]}.

\bibitem{F-3} J. Figueroa-O'Farrill, \textit{Simplicity in the Faulkner
Construction}, J. Phys. \textbf{A42}, 445206 (2009), \texttt{arXiv:0905.4900
[hep-th]}.

\bibitem{Gaiotto-Witten} D. Gaiotto and E. Witten, \textit{Janus
Configurations, Chern-Simons Couplings, And The theta-Angle in }$\mathcal{N}%
\mathit{=4}$\textit{\ Super Yang-Mills Theory}, JHEP \textbf{1006}, 097
(2010), \texttt{arXiv:0804.2907 [hep-th]}. \

\bibitem{Palmkvist-Kim} S.-S. Kim and J. Palmkvist, $\mathcal{N}\mathit{=5}$%
\textit{\ three-algebras and }$\mathit{5}$\textit{-graded Lie superalgebras}%
, J. Math. Phys. \textbf{52}, 083502 (2011), \texttt{arXiv:1010.1457 [hep-th]%
}.

\bibitem{Faulkner-Ferrar} J. R. Faulkner and J. C. Ferrar, \textit{Simple
Anti-Jordan Pairs}, Comm. Algebra \textbf{8}, no. 11, 993 (1980).

\bibitem{Exc-Reds} A. Marrani, E. Orazi, and F. Riccioni, \textit{%
Exceptional Reductions}, J. Phys. \textbf{A44}, 155207 (2011), \texttt{%
arXiv:1012.5797 [hep-th]}.

\bibitem{Baez} J. C. Baez, \textit{The Octonions}, Bull. Am. Math. Soc.
\textbf{39}, 145 (2002), \texttt{math/0105155}.

\bibitem{Yamazaki} M. Yamazaki, \textit{Octonions, }$G_{2}$\textit{\ and
generalized Lie }$3$\textit{-algebras}, Phys. Lett. \textbf{B670}, 215
(2008), \texttt{arXiv:0809.1650 [hep-th]}.

\bibitem{Castro1} C. Castro, \textit{Advances in ternary and octonionic
gauge field theories}, Int. J. Mod. Phys. \textbf{A26}, 2997 (2011).

\bibitem{Castro2} C. Castro, \textit{On Octonionic Gravity, Exceptional
Jordan Strings and Nonassociative Ternary Gauge Field Theories}, Int. J.
Geom. Meth. Mod. Phys. \textbf{9}, 1250021 (2012).

\bibitem{AGZ-rev} P. Aschieri, S. Ferrara, and B. Zumino, \textit{Duality
Rotations in Nonlinear Electrodynamics and in Extended Supergravity},
Riv.Nuovo Cim. \textbf{31} (2008), 625--708, \texttt{arXiv:0807.4039 [hep-th]%
}.

\bibitem{Cassani:2009na} D. Cassani, S. Ferrara, A. Marrani, J. F. Morales,
and H. Samtleben, \textit{A Special road to AdS vacua}, JHEP \textbf{1002}
(2010), 027, \texttt{arXiv:0911.2708 [hep-th]}.

\bibitem{Jordan:1933vh} P. Jordan, J. von Neumann, and E. P. Wigner, \textit{%
On an Algebraic generalization of the quantum mechanical formalism}, Annals
Math. \textbf{35} (1934), 29--64.

\bibitem{Cecotti:1988qn} S. Cecotti, S. Ferrara, and L. Girardello, \textit{%
Geometry of Type II Superstrings and the Moduli of Superconformal Field
Theories}, Int. J. Mod. Phys. \textbf{A4} (1989), 2475.

\bibitem{BGM} P. Breitenlohner, G. W. Gibbons and D. Maison, \textit{%
Four-Dimensional Black Holes from Kaluza-Klein Theories}, Commun. Math.
Phys. \textbf{120} (1988) 295.

\bibitem{gustavsson-2} A. Gustavsson, \textit{Selfdual strings and loop
space Nahm equations}, JHEP \textbf{0804}, 083 (2008), \texttt{%
arXiv:0802.3456 [hep-th]}.

\bibitem{Besse} A. L. Besse : \textit{\textquotedblleft Einstein Manifolds"}%
, Springer-Verlag, 1987.

\bibitem{Alekseevsky} D. V. Alekseevsky and B. N. Kimelfeld, \textit{%
Structure of Homogeneous Riemannian Spaces with zero Ricci Curvature},
Functional Anal. Appl. \textbf{9}, no. 2, 97 (1975).

\bibitem{Duff-Ferrara-gen-mirror} M. J. Duff and S. Ferrara, \textit{%
Generalized mirror symmetry and trace anomalies}, Class. Quant. Grav.
\textbf{28}, 065005 (2011), \texttt{arXiv:1009.4439 [hep-th]}.

\bibitem{Levay} P. Levay and G. Sarosi, \textit{Hitchin Functionals are
related to Measures of Entanglement}, \texttt{arXiv:1206.5066 [hep-th]}.

\bibitem{LandsbergManivel} J. M. Landsberg and L. Manivel, \textit{The
projective geometry of Freudenthal's magic square,} J. Algebra \textbf{239}
(2001), no. 2, 477--512.
\end{thebibliography}
\end{document}